\documentclass{aa}  

\usepackage{graphicx}
\usepackage{txfonts}
\usepackage{lipsum}
\usepackage{subcaption}         
\usepackage{lscape}             
\usepackage{placeins}           
\usepackage{stfloats}           
\usepackage[table]{xcolor}
\usepackage{multirow}
\usepackage[colorlinks=true, allcolors=blue]{hyperref}
\usepackage{xcolor}
\usepackage{ulem}

\begin{document}

   \title{Probing the dynamics of extreme galaxies}

   \subtitle{I. Dark matter content in ultra-diffuse galaxies}

   \author{Filippo Bouchè\inst{1,2}\fnmsep\thanks{Corresponding author: filippo.bouche-ssm@unina.it} 
        \and Chiara Buttitta\inst{3}
        \and Saboura Zamani\inst{4}
        \and Enrichetta Iodice\inst{3}
        \and Salvatore Capozziello \inst{1,2,5}
        \and Vincenzo Salzano \inst{4}
        \and Marcello Miranda\inst{1,2} 
        \and Enrico Maria Corsini \inst{6,7}
        \and Marco Mirabile \inst{8,9}
        }

   \institute{Scuola Superiore Meridionale, Via Mezzocannone 4, 80134 Napoli, Italy \and Istituto Nazionale di Fisica Nucleare, Sez. di Napoli, Via Cinthia 21, 80126 Napoli, Italy
   \and INAF $-$ Osservatorio Astronomico di Capodimonte, Salita Moiariello 16, 80131, Napoli, Italy
   \and Institute of Physics, University of Szczecin, Wielkopolska 15, 70-451 Szczecin, Poland
   \and Dipartimento di Fisica ``E. Pancini'', Università degli Studi di Napoli ``Federico II'', Via Cinthia 9, 80126 Napoli, Italy
   \and INAF $-$ Osservatorio Astronomico di Padova, Vicolo dell’Osservatorio 5, 35122 Padova, Italy
   \and Dipartimento di Fisica e Astronomia ``G. Galilei'', Universit\`a di Padova, Vicolo dell'Osservatorio 3, 35122 Padova, Italy
   \and INAF $-$ Osservatorio Astronomico d'Abruzzo, Via Maggini, 64100, Teramo, Italy
   \and Gran Sasso Science Institute, Viale Francesco Crispi 7, 67100 L'Aquila, Italy}

   \date{Received: 10 March 2026. Accepted: 8 September 2026}

  \abstract
   {We investigate the internal structure of two galaxies from the LEWIS sample: the ultra-diffuse galaxy UDG-1 and the extended dwarf galaxy LSB-6. Both systems exhibit coherent stellar rotation in combination with a non-negligible fraction of random motion, with no signs of ongoing disturbance. This work represents the first attempt to place constraints on the galaxy dynamics and on the dark matter physics of rotation-supported UDGs through the modeling of integral-field stellar kinematics.}
   {The aim of this work is to constrain the parameters that describe the internal structure of the galaxies and to assess which dark matter model and UDG formation scenario are favored by the observed stellar kinematics. We consider the standard cold dark matter paradigm as well as alternative frameworks, including fuzzy, self-interacting, and non-minimally coupled dark matter.} 
   {We modeled the galaxies as stellar spheroidal systems with typical dwarf-like thickness, embedded in spherical dark matter halos whose density profiles encode the properties of the dark matter models under consideration. We derived two-dimensional velocity fields and compared them with the observations within a Bayesian framework.} 
   {The kinematic data suggest a cuspy dark matter halo in UDG-1 and a cuspless halo in LSB-6. Beyond the standard cold dark matter scenario, all the alternative dark matter models emerge as viable candidates, although the limitations of current data do not enable us to conclusively discriminate between them. Within the fuzzy dark matter framework, we obtain mutually consistent constraints on the boson mass: $m_\alpha = 4.7^{+1.8}_{-1.1} \times 10^{-22}$ eV for UDG-1 and $m_\alpha = 6.3^{+0.4}_{-0.4} \times 10^{-22}$ eV for LSB-6. Within the self-interacting dark matter scenario, LSB-6 yields a robust constraint on the velocity-weighted cross section, $\langle \sigma v\rangle/m = 14.3^{+2.0}_{-1.9} \, \mathrm{cm}^2 \, \mathrm{km} \, \mathrm{g}^{-1} \, \mathrm{s}^{-1}$. For non-minimally coupled dark matter, we derive upper bounds on the coupling length, which implies marginal deviations from standard $\Lambda$CDM behavior. Regardless of the underlying dark matter model, both UDG-1 and LSB-6 exhibit a dark matter content comparable with the  halos of typical dwarf galaxies with similar stellar masses. Together with information from stellar kinematics, stellar population analyses, and globular clusters, these results support a scenario in which UDG-1 and LSB-6 originate from puffed-up dwarfs.}
   {By probing the kinematics of extremely low-surface-brightness systems, we infer their internal structure and translate these results into constraints on dark matter particle properties. Our analysis also provides new insights into the formation mechanisms of UDGs. This work is intended as the first in a series exploiting kinematic data of extremely low-surface-brightness galaxies to test fundamental physics, from dark matter to modified theories of gravity.} 


   \maketitle
   \nolinenumbers

\section{Introduction}

In the standard model of galaxy formation, dark matter (DM) halos provide the gravitational potential wells within which baryonic matter condenses and forms galactic structures from small to cluster scale, through a hierarchical process that traces the filamentary DM structure of the Universe \citep{PressSchechter1974, Binney1977, WhiteRees1978}. The current cosmological standard model, $\Lambda$CDM, is built upon the assumptions that collisionless, cold dark matter (CDM) dominates the matter budget of the Universe, and a cosmological constant ($\Lambda$) accounts for the accelerated cosmic expansion \citep{PlanckCollab2020}. Despite its remarkable success on large scales, the $\Lambda$CDM model faces persistent challenges at galaxy scales \citep[see][and references therein]{Bullock:2017xww, Sales:2022ich}. These tensions are most pronounced in the low-mass regime, where baryons contribute little to the gravitational potential and DM governs the dynamics. The $\Lambda$CDM model predicts a large population of low-mass halos ($M_{h}\sim10^{5-9}\,M_\odot$), whose luminous counterparts should be observable in the form of dwarf galaxies and provide a benchmark for testing cosmological models. This regime is highly relevant for alternative models, such as non-standard DM candidates or modified gravity, proposed to address small-scale tensions \citep{deMartino:2020gfi}.

In the realm of dwarf galaxies, ultra-diffuse galaxies (UDGs) play a special role. First empirically defined in 2015, these dwarf-like systems have stellar masses of $M_\ast\sim10^{7.5-9} \,M_\odot$ and are characterized by large effective radii ($R_{\rm eff}>1.5\,{\rm kpc}$) and extremely low central surface brightnesses ($\mu_{0,g}>24\,{\rm mag\,arcsec^{-2}}$), which places them at the low-luminosity tail of the dwarf galaxy size-luminosity distribution \citep{vanDokkum2015, Lim2020}. Advances in low-surface brightness (LSB) surveys have since enabled the identification of large UDG samples across a wide range of environments \citep{Lim2020, Iodice_2020, Marleau2021, LaMarca2022b, Zaritsky2022}, which has revealed a broad diversity in their structural and dynamical properties \citep[see][and references therein]{Buzzo2025b}. In particular, their DM content remains highly debated, with studies reporting systems that are DM-dominated \citep{vanDokkum:2019fdc, Gannon2021, Gannon2024, Iodice_2023, Ferre-Mateu2023, Mirabile2026}, consistent with dwarf-sized halos, or nearly DM-free \citep{vanDokkum2018, vanDokkum2019, Mancera-Pina2022}. These extreme cases make UDGs valuable laboratories for probing the interplay between baryons and DM, as well as for testing alternatives to General Relativity \citep{Capozziello:2006ph, Capozziello:2008ny, Capozziello:2011et,  Cardone:2011ze, CANTATA:2021asi}. This potential has been recently illustrated by the study of the gas-rich UDG AGC\,114905, whose extended \ion{H}{I} rotation curve exhibits an exceptionally low amplitude ($\Delta V \sim 23\,\mathrm{km\,s^{-1}}$). By modeling this system, \cite{Mancera-Pina2024} tested standard CDM, fuzzy dark matter (FDM), and self-interacting dark matter (SIDM) scenarios. Similarly, a wide range of alternative frameworks -- including modified gravity theories \citep{Laudato:2022vmq, Laudato:2022drz, Benetti:2023imt, Bhatia:2023pts, Bouche:2024qhy} and non-standard DM models \citep{Wasserman_2019, Yang:2020iya, Pozo:2020fft, Zamani:2026cxk} -- have been extensively applied to well-known UDGs such as NGC~1052-DF2 and NGC~1052-DF4, which are characterized by extremely low values of velocity dispersion, and Dragonfly~44.

Assessing the amount of DM is essential to identify the physical processes that acted on the galaxy, affecting its structure and transforming it into a UDG. Several theories have been invoked to explain the plethora of UDG properties. In the failed galaxy scenario \citep{vanDokkum2015}, UDGs originate from bright, high-luminosity galaxies that lost their gas at early times, becoming red, quenched, and DM-dominated. Alternatively, UDGs may form from puffed-up dwarfs, whose stellar distributions are inflated by internal processes -- such as strong supernova feedback or high halo spin \citep{Amorisco2016, Cintio2017, Rong2017} -- or by environmental interactions \citep{Yozin2015, Tremmel2020}. These UDGs have properties similar to those of the parent dwarf galaxy and live in dwarf-like DM halos. Finally, blue, gas-rich, DM-free UDGs may originate from collisional debris from galaxy mergers \citep{Lelli2015, Duc2015}, or ram-pressure-stripped gas clumps \citep{Poggianti2019}, while red, quenched, DM-free systems can arise from gravitationally bound, collisional debris from high-velocity galaxy encounters \citep{Shin2020, vanDokkum2022}.

In addition to formation scenarios, investigating the halo properties of UDGs provides a unique opportunity to probe the fundamental nature of DM. Given the broad diversity of their structural parameters and environmental conditions, UDGs constitute a powerful laboratory to test the standard CDM paradigm and its alternatives. Such constraints can be combined with independent astrophysical and cosmological probes to build a coherent, data-driven picture of the dominant matter component of the Universe. Current observations constrain viable DM candidates to be non-relativistic and weakly interacting \citep{Bertone:2004pz, Irsic:2023equ}. However, the persistent absence of direct detection \citep{Liu:2017drf} leaves room for a broad spectrum of theoretical possibilities, including weakly interacting massive particles (WIMPs, \citealp{Chang:2013oia}), axion-like particles (ALPs, \citealp{Marsh:2015xka, Addazi:2024mii, Addazi:2026kxf}), SIDM, \citep{Spergel:1999mh}, and non-minimally coupled (NMC) scenarios \citep{Bruneton:2008fk}. The microphysical properties of these candidates imprint distinct signatures on DM halo structure and galaxy dynamics. In this work, we focus on the four aforementioned frameworks; we derive constraints on their key parameters and explore the implications for UDG halo properties and formation scenarios.

The paper is organized as follows. In Sec.~\ref{sec:kinematic_data}, we introduce the galaxy sample and describe the kinematic data employed in our analysis. The DM models considered in this study are presented in Sec.~\ref{sec:dark_matter}, with particular emphasis on their astrophysical implications. In Sec.~\ref{sec:galaxy_modeling}, we present the assumptions underlying our analysis along with the dynamical models adopted for the stellar and DM components to predict the observed kinematics. Section~\ref{sec:statistical_analysis} provides an overview of our statistical methodology, including the definition of the likelihood and priors, the sampling strategy, and the Bayesian tools used for model comparison. The results of our analysis are presented in Sec.~\ref{sec:results}, where we discuss the constraints on DM properties and halo profiles. The implications for UDG formation mechanisms are discussed in Sec.~\ref{sec:formation_UDGs}. Finally, our conclusions are summarized in Sec.~\ref{sec:conclusions}. Additional DM halo models, together with their corresponding kinematic constraints, are presented in Appendix~\ref{Appendix_CDM}. The posterior distributions and the normalized residual maps from our analysis are shown in Appendices~\ref{AppendixB:fig_posteriors} and~\ref{AppendixC:residuals}, respectively.

\section{Galaxy sample}\label{sec:kinematic_data}

The dataset adopted in this work belongs to the LEWIS project (Looking into the faintEst WIth MUSE, Prog. Id. 108.222P). This is an ESO Large Program, awarded with more than 130 hours with the integral-field (IF) Multi Unit Spectroscopic Explorer (MUSE) mounted on ESO's Very Large Telescope (VLT) in Chile. The LEWIS project mapped 30 UDGs in the Hydra I cluster of galaxies \citep{Iodice_2023}. Thanks to the IF nature of the data, it has been possible to retrieve both integrated and spatially resolved stellar kinematics \citep{Buttitta_2025} as well as global stellar populations properties \citep{Doll2026}.

Galaxies in the LEWIS sample are characterized by low values of stellar velocity dispersions, $\sigma_{\rm eff} \sim 20-35$ km s$^{-1}$, measured by co-adding the spaxels within an aperture of an effective radius ($R_{\rm eff}$) and performing spectral fitting on the stacked spectrum. The analysis of the stellar velocity fields has been successfully performed for 18 of the 30 LEWIS galaxies \citep{Buttitta_2025}. Seven UDGs show clear hints of stellar rotation, with a rotation curve velocity amplitude of $\Delta V \sim 25-50$ km s$^{-1}$, five do not rotate, whereas the remaining six galaxies have unconstrained kinematic patterns. For five out of the seven rotating UDGs it was possible to derive the dynamical support; these galaxies turned out to be rotation-supported ($\lambda_R\sim0.3$, \citealp{Emsellem_2011}). These findings suggest that the dynamics of these UDGs comprise both random and ordered motion components, with the ordered component contributing most.

From this subsample of five targets, we excluded three galaxies due to potential complications in the dynamical modeling. UDG-8 exhibits an off-centered, elongated bar-like structure in its inner regions, introducing significant non-axisymmetric components. LSB-7 and LSB-8 display nearly round morphologies ($\epsilon \lesssim 0.2$, with $\epsilon$ the ellipticity of the isophote) and show pronounced velocity gradients misaligned with their photometric major axes, likely driven by external tidal distortions rather than internal stellar kinematics. As a result, only two systems remain suitable for our analysis: UDG-1 and LSB-6.

UDG-1 is a genuine UDG according to the definition of \citealt{vanDokkum2015} with $\mu_{0,g} = 24.2 \pm 0.1$ mag arcsec$^{-2}$ and $R_{\rm eff} = 1.75 \pm 0.12$ kpc \citep{Iodice_2023}. The isophotal analysis indicates a mildly elliptical shape ($\epsilon \sim 0.25$) and a low central concentration of light, with an azimuthally averaged profile shallower than exponential (Sérsic $n = 0.62$). UDG-1 is located in the innermost region of the Hydra I cluster, classified as a very early infaller \citep{Buttitta_2025}. The analysis of MUSE data shows that the galaxy exhibits a low stellar velocity dispersion ($\sigma_{\rm eff} \sim 23$ km s$^{-1}$) and modest rotation along the major axis ($\Delta V \sim 44$ km s$^{-1}$), has a high DM content ($M_{\rm dyn} \sim 2.3 \cdot 10^9 M_\odot$) with a dynamical mass-to-light ratio of $\Upsilon_{\rm dyn} \sim 40$. 

LSB-6 exhibits different morphological and structural properties. With a large effective radius ($R_{\rm eff}\,=\,4~\,\pm~1$ kpc) and relatively bright central surface brightness ($\mu_{0,g} = 23.0 \pm 0.2$ mag arcsec$^{-2}$), it is classified as an extended dwarf \citep{Buttitta_2025}. Isophotal analysis reveals elongated ellipses ($\epsilon \sim 0.5$), which are boxier in the inner regions and slightly tilted and disky in the outskirts \citep{Buttitta_2025}. The light distribution is more centrally concentrated, with a profile steeper than exponential (Sérsic $n = 1.62$, \citealp{Iodice_2020}). LSB-6 resides at large clustercentric distances and is thus classified as a late infaller \citep{Buttitta_2025}. It exhibits low stellar velocity dispersion ($\sigma_{\rm eff} \sim 28$ km s$^{-1}$) with mean stellar rotation of $\Delta V \sim 32$ km s$^{-1}$, and contains a large amount of DM ($M_{\rm dyn} \sim 2.4 \cdot 10^9 M_\odot$) with a dynamical mass-to-light ratio of $\Upsilon_{\rm dyn} \sim 50$. 

The quoted values for $M_{\rm dyn}$ have been calculated by using the luminosity-weighted second velocity moment $\langle v_{\rm rms}^2\rangle = \langle v_{\rm los}^2 + \sigma_{\rm los}^2\rangle$ computed in an aperture equal to $R_{\rm eff}$ and adopting the formula reported in \cite{Wolf2010} to account for both rotation and pressure-supported components (see \citealt{Buttitta_2025} for details). The values of the dynamical mass presented here and computed in the forthcoming analysis thus correspond to the mass enclosed within the half-light radius of the galaxy.

\section{Dark matter candidates}\label{sec:dark_matter}

Dark matter is a fundamental component of modern cosmological models, shaping the clustering properties of the Universe. Evidence for its presence spans a wide range of scales, from temperature anisotropies in the cosmic microwave background to galaxy rotation curves, the filamentary large-scale structure, and gravitational lensing phenomena \citep{Bertone:2004pz}. However, DM has so far been detected only via its gravitational imprint, as all experimental efforts to identify the underlying fundamental particles have failed \citep{PICO:2019vsc, DarkSide:2022dhx, ATLAS:2024kpy, CMS:2024zqs, LZ:2024zvo, PandaX:2024qfu, XENON:2025vwd}. This lack of direct detection leaves room for a plethora of theoretical scenarios, encompassing both particle candidates beyond the Standard Model and modifications of General Relativity. 

In this work, we focus on the former class of models, exploring how different particle candidates explain the observed kinematics of UDG-1 and LSB-6. We thus aim to constrain the parameters that characterize their internal structure and to assess which DM model and formation scenario are preferred by the data. Below, we briefly review the DM models under investigation, highlighting their main astrophysical properties.

\subsection{Standard cold dark matter}\label{sec:standardCDM}

Cold DM, composed of non-relativistic particles, represents the standard paradigm within the $\Lambda$CDM model. Observations strongly support this framework, placing tight constraints on any warm or mildly relativistic DM component \citep{Enzi:2020ieg, Irsic:2023equ}. Over the years, several CDM candidates have been proposed. Among them, WIMPs, especially those arising in supersymmetric extensions of the Standard Model, have long been regarded as the most promising candidates. Their thermal freeze-out in the early Universe would naturally yield a relic abundance consistent with the present-day DM density \citep{Chang:2013oia}. Other viable CDM candidates include Quantum Chromodynamics axions and ALPs \citep{Marsh:2015xka}, which are non-relativistic despite their low masses, as they would be produced with very small initial momenta in the early Universe. Heavy sterile neutrinos, which offer a natural explanation for the neutrino masses through the seesaw mechanism, have also been considered \citep{Boyarsky:2018tvu}, as well as primordial black holes as macroscopic contributors to the CDM density \citep{Carr:2021bzv}.

When implemented in numerical simulations, the CDM paradigm has proven remarkably successful in reproducing a wide range of astrophysical and cosmological observables \citep{Angulo:2021kes}. In particular, N-body simulations predict the emergence of a nearly universal DM density profile across halo mass scales, commonly described by the Navarro-Frenk-White (NFW) form \citep{Navarro:1995iw},
\begin{equation}\label{NFW}
\rho_{\mathrm{NFW}}(r) = \rho_s \, \bigg(\frac{r}{r_s}\bigg)^{-1} \bigg( 1 + \frac{r}{r_s} \bigg)^{-2} \, .
\end{equation}
Here, $\rho_s$ is the characteristic halo density and $r_s$ is the scale radius at which the logarithmic slope of the density distribution satisfies $\mathrm{d}\ln \rho(r)/\mathrm{d}\ln r = -2$, hence denoted as $r_s = r_{-2}$. The dimensionless concentration parameter is then defined as
\begin{equation}\label{eq:c200}
c_{200} = \frac{r_{200}}{r_{-2}} \, ,
\end{equation}
where $r_{200}$ is the radius within which the mean enclosed density is 200 times the critical density $\rho_c$ of the Universe at the halo redshift. The corresponding characteristic halo density reads
\begin{equation}
\rho_s = \frac{200}{3} \rho_c \, c_{200}^3 \left[ \ln(1 + c_{200}) - \frac{c_{200}}{1 + c_{200}} \right]^{-1} \, .
\end{equation}
In this work, we adopted $\{c_{200}, \, \log_{10}M_{200}\}$ as the free parameters of the NFW model, where $M_{200}$ denotes the total mass enclosed within the overdensity radius $r_{200}$.

Despite its overall success, the $\Lambda$CDM model exhibits several tensions on sub-Megaparsec scales \citep{Bullock:2017xww}. CDM-only simulations predict a nearly universal DM halo structure characterized by a steep central cusp (Eq.~\ref{NFW}), in contrast to observations of many dwarf galaxies that instead favor cored density profiles with significantly lower central densities \citep{2017MNRAS.467.2019R, 2018MNRAS.480..927P} -- the so-called cusp-core problem. This tension may partly arise from modeling uncertainties or observational degeneracies \citep{2009MNRAS.393L..50E, 2023MNRAS.521.1316R}, but it can also be explained by physical processes such as baryonic feedback, which induce fluctuations in the gravitational potential that heat the orbits of DM particles \citep{Sales:2022ich, Boldrini2026}. However, this mechanism struggles to account for the large diversity observed in the inner slopes of DM density profiles and in the corresponding rotation curves.

To capture this diversity, we also considered the cored profile proposed by \cite{Burkert:1995yz}, which successfully describes both dwarf galaxies \citep{Burkert:1997fz} and generic disk galaxies \citep{Salucci:2000ps}. The Burkert profile reads
\begin{equation}\label{Burkert}
    \rho_{\mathrm{B}}(r) = \rho_s \, \bigg(1 + \frac{r}{r_s}\bigg)^{-1} \bigg( 1 + \frac{r^2}{r_s^2} \bigg)^{-1} \, .
\end{equation}
In this case, the scale radius satisfies the relation $r_{-2} \approx 1.521 \, r_s$, and the concentration parameter is defined as in Eq.~\ref{eq:c200}. The characteristic density can be written as
\begin{equation}
    \rho_s = \frac{200}{3} \rho_c \, \frac{\left(1.521 \, c_{200} \right)^3}{\frac{\log\left\{ \left[ 1 + (1.521 \, c_{200})^2 \right] \, (1 + 1.521 \, c_{200})^2 \right\}}{4} - \frac{\arctan(1.521 \, c_{200})}{2}} \, .
\end{equation}
As in the NFW case, we adopted $\{c_{200}, \, \log_{10}M_{200}\}$ as the free parameters of the Burkert profile.

Finally, we performed our dynamical analysis in the standard $\Lambda$CDM framework using the Einasto profile \citep{Einasto_1965},
\begin{equation}\label{Einasto}
    \rho_{E}(r) = \rho_s \, \mathrm{exp} \left\{ -\frac{2}{\gamma} \left[ \left( \frac{r}{r_s} \right)^\gamma -1 \right] \right\} \, ,
\end{equation}
where the concentration parameter is defined as in Eq.~\ref{eq:c200}, with $r_{-2} = r_s$, and the characteristic density reads
\begin{equation}
    \rho_s = \frac{200}{3} \rho_c \, c_{200}^3 \gamma \, \frac{\exp \left( - \frac{2}{\gamma} \right) \left( \frac{2}{\gamma} \right)^{3/\gamma}}{\Gamma \left( \frac{3}{\gamma} \right) - \Gamma \left( \frac{3}{\gamma}, \frac{2}{\gamma} \, c_{200}^\gamma \right)}  \, .
\end{equation}
The Einasto profile is cuspless, featuring a finite central density and a logarithmic slope that gradually approaches zero, without developing the extended flat core characteristic of the Burkert profile. Within this framework, larger values of the shape parameter $\gamma$ correspond to increasingly cored density distributions. Equation~\ref{Einasto} shares the same analytic form as the Sérsic profile \citep{1968adga.book.S, Graham:2005fy} -- commonly used for the distribution of stars -- although it describes the three-dimensional mass density rather than the two-dimensional light distribution. The Einasto profile has been shown to provide an excellent description of $\Lambda$CDM halos in numerical simulations across a wide range of masses, from subhalos hosting satellite dwarf galaxies to galaxy clusters \citep{Graham:2005xx, Einasto_2012, Di_Cintio_2013}, naturally accounting for the mass dependence of halo shapes.

Additional density profiles have been explored to encompass the full diversity of halo shapes inferred from observations. A detailed discussion of these models is provided in Appendix~\ref{Appendix_CDM}.

\subsection{Fuzzy dark matter}

Still within the CDM framework, FDM offers a viable alternative to the particle candidates discussed in the previous section. FDM consists of ultra-light ALPs, i.e., bosons with masses on the order of $10^{-22}$ eV. Such particles arise naturally in the string theory framework and can be produced through the misalignment mechanism \citep{Hui:2021tkt}. The resulting FDM population would be intrinsically non-relativistic, and its relic abundance would read
\begin{equation}
    \Omega_{\mathrm{ALPs}} \sim 0.1 \left(\frac{f}{10^{17} \, \mathrm{GeV}} \right)^2 \left(\frac{m_\alpha}{10^{-22} \, \mathrm{eV}} \right)^{1/2} \, ,
\end{equation}
where $f$ denotes the ALP decay constant — an energy scale typically close to the Planck scale — and $m_\alpha$ is the ALP mass. For sub-Planckian values of the decay constant, $f<4\times 10^{17}$ GeV, the FDM relic abundance matches the observed CDM density today for ALP masses $m_\alpha>10^{-24}$ eV \citep{Marsh:2015xka}. Moreover, since ALP interactions scale as $\sim 1/f^{\,n}$ with $n$ a positive integer, they are extremely suppressed, making FDM effectively collisionless on cosmological scales.

Fuzzy dark matter exhibits a richer phenomenology compared to standard CDM, giving rise to distinctive astrophysical and cosmological signatures that may be observable. In particular, ultra-light ALPs have a macroscopic de Broglie wavelength that exceeds the mean interparticle separation in galactic environments. As a result, FDM behaves collectively as a classical wave and can form coherent standing-wave configurations at the centers of gravitationally bound systems, giving rise to dense central cores within DM halos \citep{Schive:2014dra}. These cores correspond to solitonic solutions of the Schr\"odinger-Poisson equations, which arise when the quantum pressure associated with the uncertainty principle counterbalances gravitational attraction. The soliton density profile can be written as in \cite{Marsh:2015wka}
\begin{equation}
    \rho_{\mathrm{sol}}(r) = \rho_s\left[ 1 + \left( \frac{r}{r_{\mathrm{sol}}} \right)^2 \right]^{-8} \, ,
\end{equation}
where the soliton radius is inversely proportional to the ALP mass $m_\alpha$,
\begin{equation}\label{eq:rsol}
    r_{\mathrm{sol}} = \frac{1}{\alpha m_\alpha} \, ,
\end{equation}
and the corresponding soliton density reads
\begin{equation}\label{eq:rhosol}
    \rho_s = \rho_c \left( \frac{5 \times 10^4}{\alpha^4} \right) \left( \frac{h}{0.7} \right)^{-2} \left( \frac{m_\alpha}{10^{-22} \, \mathrm{eV}} \right)^{-2} \left( \frac{r_{\mathrm{sol}}}{\mathrm{kpc}} \right)^{-4} \, .
\end{equation}
Here, $\alpha = 0.230$ is obtained from numerical solutions of the Schr\"odinger-Poisson system \citep{Marsh:2015wka}, and $h$ denotes the dimensionless Hubble parameter. The inner solitonic core exhibits a sharp boundary, beyond which the density of the host halo rapidly dominates. The resulting FDM density profile of a galactic halo can therefore be written as
\begin{equation}\label{eq:fuzzyDM}
    \rho_{\mathrm{FDM}}(r) = 
    \begin{cases}
          \rho_{\mathrm{sol}}(r) & \mathrm{if}\; r \leq r_{t} \; , \\
          \rho_{\mathrm{NFW}}(r) & \mathrm{if }\; r > r_{t} \; ,
    \end{cases}
\end{equation}
where $r_t$ denotes the transition radius at which the soliton and NFW densities coincide. Accordingly, we adopted $\{c_{200}, \, m_\alpha, \, r_t\}$ as the free parameters of the FDM profile. Using Eqs.~\ref{eq:rsol}$\,$-$\,$\ref{eq:rhosol} together with the continuity condition $\rho_{\mathrm{sol}}(r_t)=\rho_{\mathrm{NFW}}(r_t)$, both $M_{200}$ and $r_{\mathrm{sol}}$ can be expressed in terms of these three parameters.

\subsection{Self-interacting dark matter}

Another possible fundamental solution to the small-scale challenges of the $\Lambda$CDM model is provided by SIDM \citep{Spergel:1999mh}. In this framework, DM particles undergo elastic $2\rightarrow2$ scattering, enabling heat transfer from the hotter outer halo to the cooler inner regions and driving the core toward thermalization. This process naturally generates cored density profiles, while preserving the outer halo structure where the lower DM density suppresses self-interactions. Moreover, since halos were less dense in the early Universe, structure formation remains unaffected, provided that the self-interaction cross section is sufficiently small. N-body simulations indicate that a self-interaction cross section of order $\sigma/m \sim 0.5 \; \mathrm{cm}^2\,\mathrm{g}^{-1}$ can alleviate small-scale tensions while remaining consistent with astrophysical and cosmological constraints. However, a velocity-dependent cross section, typically $\sigma \propto 1/v$, is required to satisfy observational bounds across different environments \citep{Tulin:2017ara}. Such values cannot be achieved within the standard WIMP framework, where weak-scale interactions imply $(\sigma/m)_{\mathrm{WIMPs}} \ll 0.5 \; \mathrm{cm}^2\,\mathrm{g}^{-1}$. The required magnitude is similar to that of nuclear interactions mediated by pions, although SIDM self-interactions cannot be strongly coupled.

Beyond numerical simulations, semi-analytical models can be used to study the impact of SIDM on halos \citep{Kaplinghat:2015aga}. In the collisional regime, the time-independent Jeans equation, describing hydrostatic equilibrium in a self-gravitating fluid, can be coupled to the Poisson equation,
\begin{equation}
    \sigma_0^2 \nabla^2 \ln \rho_{\mathrm{dm}} = -4 \pi G \left( \rho_{\mathrm{dm}} + \rho_{\mathrm{b}}\right) \, ,
\end{equation}
where $\sigma_0$ is the isotropic and isothermal velocity dispersion, $\rho_{\mathrm{dm}}$ is the SIDM density, and $\rho_{\mathrm{b}}$ denotes the baryon density. The solution to this equation is a pseudo-isothermal density profile,
\begin{equation}
    \rho_{\mathrm{iso}} (r) = \rho_0 \left[1 + \left( \frac{r}{r_c} \right)^2 \right]^{-1} \, ,
\end{equation}
where $r_c$ is the scale radius and $\rho_0$ is the characteristic density,
\begin{equation}
    \rho_0 = \frac{200}{3} \rho_c \, \frac{(c_{200}^{\mathrm{iso}})^3}{c_{200}^{\mathrm{iso}} - \arctan\left(c_{200}^{\mathrm{iso}}\right)} \, .
\end{equation}
The collisional description applies only to the inner halo, where the DM density is sufficiently high for self-interactions to drive thermalization. At larger radii, the density drops and scattering becomes inefficient, so the system transitions to the collisionless regime. The resulting SIDM density profile is therefore
\begin{equation}
    \rho_{\mathrm{SIDM}} (r) = 
    \begin{cases}
          \rho_{\mathrm{iso}} (r) & \mathrm{if}\; r \leq r_{t} \; , \\
          \rho_{\mathrm{NFW}}(r) & \mathrm{if }\; r > r_{t} \; ,
    \end{cases}
\end{equation}
where $r_t$ is the transition radius at which the collisional and collisionless solutions match. In this work, we adopted $\{c_{200}, \, c_{200}^{\mathrm{iso}}, \, \log_{10}M_{200}^{\mathrm{iso}}, \, r_t \}$ as the free parameters of the SIDM profile, while the total halo mass $M_{200}$ is determined by imposing the continuity condition $\rho_{\mathrm{iso}}(r_t)~=~\rho_{\mathrm{NFW}}(r_t)$.

The self-interaction cross section can be inferred from the fitted SIDM profile by requiring that the average scattering rate per particle, multiplied by the halo age, equals unity at $r_t$ \citep{Kaplinghat:2015aga}. This yields
\begin{equation}\label{eq:SIDM_cross_section}
    \frac{\langle \sigma v \rangle}{m} \approx \frac{1}{\rho_{\mathrm{SIDM}}(r_t) \, t_{\mathrm{halo}}} \, ,
\end{equation}
where $\sigma$ denotes the scattering cross section, $m$ the DM particle mass, and $v$ the relative velocity between particles. The latter can be written as in \cite{Yang:2023jwn}
\begin{equation}\label{eq:SIDM_velocity}
    v \simeq 1.065 \, r_s \sqrt{G_{\mathrm{N}} \, \rho_s} \, ,
\end{equation}
where $G_{\mathrm{N}}$ is the gravitational constant and $(r_s, \,\rho_s)$ the NFW scale parameters. Assuming the fiducial $\Lambda$CDM cosmology, the halo age $t_{\mathrm{halo}}$ in Eq.~\ref{eq:SIDM_cross_section} is estimated from \cite{Correa:2014xma}
\begin{equation}
    \begin{aligned}
        z_{\mathrm{halo}} = &-0.0064 \left[ \log_{10}\left( \frac{M_{200}}{10^{10} M_\odot} \right) \right]^2 \\
        &- 0.1043 \, \log_{10}\left( \frac{M_{200}}{10^{10} M_\odot} \right) + 1.4807 \, .
    \end{aligned}
\end{equation}

\subsection{Non-minimally coupled dark matter}

An alternative to DM particle models that modify the halo density profile is provided by scenarios in which the halo structure remains unchanged but the DM gravitational interaction is non-standard. In this context, models in which DM couples non-minimally to gravity were first proposed by \cite{Bruneton:2008fk}. The NMC approach treats DM as a macroscopic fluid, without assuming any specific particle candidate, and is therefore independent of its microscopic nature. When the mean free path of the DM fluid is comparable to or exceeds the scale over which spacetime curvature varies, a NMC between DM and geometry can naturally arise. This effect should not be interpreted as the introduction of a new fundamental force or additional degrees of freedom, as it emerges as an effective behavior after averaging over small-scale structure, thereby avoiding stringent early Universe constraints. The NMC between DM and gravity thus affects both the evolution of the DM fluid and its role as a source of the gravitational field, potentially modifying the internal kinematics of galaxies.

Within this framework, the Einstein-Hilbert action is extended by an additional NMC term \citep{Bettoni:2015wla,Zamani:2024qbx}, which can take either a conformal form,
\begin{equation}\label{eq:conformal_action_NMC}
S_{\rm NMC}^{\rm (c)}
= \frac{1}{16 \pi G_{\mathrm{N}}}
\int d^{4}x \sqrt{-g}\,\Big[\alpha_c F_c(\rho)R\Big] \, ,
\end{equation}
or a disformal one,
\begin{equation}\label{eq:disformal_action_NMC}
S_{\rm NMC}^{\rm (d)}
= \frac{1}{16 \pi G_{\mathrm{N}}}
\int d^{4}x \sqrt{-g}\,\Big[\alpha_d F_d(\rho)R_{\mu\nu}u^{\mu}u^{\nu}\Big] \, .
\end{equation}
Here, $R$ and $R_{\mu\nu}$ denote the Ricci scalar and tensor, respectively, describing spacetime curvature, while $u^{\mu}$ is the fluid four-velocity. The parameters $\alpha_c$ and $\alpha_d$ control the strength of the non-minimal interaction, whereas $F_c(\rho)$ and $F_d(\rho)$ are arbitrary functions of the density. To avoid unnecessary complexity and confine the effect to the dark sector, we followed \cite{Bettoni:2011fs} and adopted $F_{c,d}(\rho)=\rho_{\rm DM}$. With this choice, the non-minimal coupling involves only the DM component, while baryons remain minimally coupled and behave as in the standard model. 

Since our goal is to test galaxy kinematics, we focused on the Newtonian limit of the NMC theory. At this level, conformal and disformal couplings lead to the same modified Poisson equation, differing essentially only by a rescaling of the coupling length. The Newtonian limit yields
\begin{equation}\label{eq:poissonNMC}
    \nabla^2\Phi(r) =4 \pi G_{\mathrm{N}} \left[\rho(r) - \epsilon L^2 \nabla^2 \rho_{DM}(r) \right] \, ,
\end{equation}
where $\alpha_{c,d}=\epsilon L^2$, with $L^2=2\ell^2$ in the conformal case and $L^2=\ell^2$ in the disformal one. Here, $\epsilon = \pm 1$ denotes the polarity of the coupling and $\ell$ the characteristic coupling length of the NMC model. $\Phi(r)$ is the gravitational potential, $\rho(r)$ is the total matter density and $\rho_{\rm DM}(r)$ is the DM density. Since the framework does not assume a specific microscopic DM candidate, we explored the NMC scenario using the three density profiles investigated within the standard CDM framework (Sec.~\ref{sec:standardCDM}). The free parameters of the DM model are therefore $\{\theta_{\mathrm{CDM}}, \, \log_{10}L \}$, where $\theta_{\mathrm{CDM}}$ denotes the parameter of the standard CDM profile.

\section{Galaxy kinematics}\label{sec:galaxy_modeling}

Modeling the internal dynamics of UDGs is challenging, and only a limited number of studies have addressed this problem using stellar or gas kinematics. \cite{ManceraPina2020} investigated the internal structure of several samples of gas-rich UDGs in low-density environments and in an isolated field using resolved \ion{H}{I} kinematics. These systems typically exhibit gas velocity dispersions of $\sigma \sim 5-8$ km s$^{-1}$ and maximum rotation velocities of $v_{\rm max} \sim 20-40$ km s$^{-1}$. The \ion{H}{I} data extend out to several kpc ($\sim5-8\,R_{\rm eff}$), thus motivating the use of tilted-ring and thin-disk assumptions in the dynamical modeling ($v_{\rm max}/\sigma \sim 4-5$).

Obtaining deep optical spectroscopy of UDGs is observationally expensive, and accurate stellar kinematics measurements require high-quality spectra. As a result, only two studies have so far derived spatially resolved stellar kinematics and attempted to reconstruct the internal structure of UDGs using IF spectroscopy of stellar tracers. \cite{Emsellem2019} analyzed MUSE observations of the well-known UDG NGC1052-DF2 \citep{vanDokkum2018}. They measured a low stellar velocity dispersion of $\sigma_{\rm eff} \sim 10$ km s$^{-1}$ and detected a weak velocity gradient of $\sim6$ km s$^{-1}$ along the minor axis of the galaxy. Given the extension of the kinematic map ($\sim1.5\,R_{\rm eff}$), these results, together with the roundish shape of the galaxy ($q=0.85$), indicate a system with comparable ordered and random motions ($v/\sigma_{\rm eff} \lesssim 1$), resembling a thick stellar body.

More recently, \cite{Buzzo2025} presented a detailed spectroscopic analysis of the peculiar ionized-gas-rich UDG GAMA\,526784, characterized by an old central stellar body with coherent rotation and star-forming outskirts. The authors report a stellar velocity dispersion of $\sigma_{\rm eff,\ast} \sim 10$ km s$^{-1}$ and a higher value for the gas component, $\sigma_{\rm eff,gas} \sim 30$ km s$^{-1}$. Both components exhibit coherent rotation: the stellar component reaches $v_{\rm max,\ast} \sim 18$ km s$^{-1}$ at $\sim0.8-1\,R_{\rm eff}$, while the ionized gas attains higher velocities, $v_{\rm max,gas} \sim 50$ km s$^{-1}$ at $\sim1.5-2\,R_{\rm eff}$. The gas rotation is misaligned by $\sim20^\circ$ with respect to the stellar motion, suggesting a recent interaction or external perturbation. These results indicate that both stellar and gas components display a combination of ordered and random motions, with rotation moderately dominating over dispersion ($v_{\rm max,\ast}/\sigma_{\rm eff,\ast} \sim v_{\rm max,gas}/\sigma_{\rm eff,gas} \sim 1.6-1.8$).

In this work, we attempted to constrain the internal structure of two UDGs in the Hydra I cluster using spatially resolved stellar kinematics obtained with MUSE, as part of the LEWIS project. Since spatial coverage is crucial not only for providing stronger constraints on the dynamical modeling but also for obtaining a reliable assessment of the dynamical support of the galaxies, we reanalyzed the MUSE datacubes of both galaxies to extend the radial coverage of the velocity field. We adopted the same strategy described in \cite{Buttitta_2025} and derived updated estimates of the rotation velocity at larger radii. For UDG-1, we were able to extend the velocity profile out to $R_{\rm max}\sim1.5\,R_{\rm eff}$, reaching a local maximum velocity of $v_{R_{\rm max}}\sim55$ km s$^{-1}$. Similarly, for LSB-6 we measured $v_{R_{\rm max}}\sim45$ km s$^{-1}$ at $R_{\rm max}\sim0.9\,R_{\rm eff}$. For comparison with literature data, we obtained a projected value of $v_{\rm R_{max}}/\sigma_{\rm eff}\sim2.4$ for UDG-1 and $v_{\rm R_{max}}/\sigma_{\rm eff}\sim1.6$ for LSB-6, respectively.

Consistently with previous results obtained in \cite{Buttitta_2025}, the two galaxies are thus supported by rotation ($v/\sigma>1$), but they also contain a non-negligible fraction of random motions. The morphology provides additional insight into the dynamical structure: the light distribution of UDG-1 reveals concentric elliptical isophotes ($ q \sim 0.7$) and a surface-brightness profile with a Sérsic index of $n = 0.62$. LSB-6, instead, exhibits a steeper light profile ($n=1.6$) and more elongated isophotes ($q\sim0.5$), boxier in the inner regions, and more disky and tilted in the outer regions. These morphological signatures, together with their stellar kinematic information, suggest that both systems resemble a thick, mildly flattened stellar body rather than classical dwarf spheroidals, which typically show dispersion-dominated kinematics, or gas-rich irregular UDGs, characterized by a cold, thin, rotating component.

Our dynamical model, under reasonable approximation, aims to reconstruct the internal structure of these systems. In forthcoming sections, we present the formalism adopted to model the stellar and DM contributions in UDG-1 and LSB-6. We neglected any other contributions from the baryonic component. Despite Hydra I having been covered by the WALLABY@ASKAP survey \citep{Westmeier2022}, the \ion{H}{I} mass sensitivity at 5$\sigma$ only allows for the detection of $M_{\rm HI} \gtrsim 10^8 M_\odot$. We therefore cannot exclude the presence of \ion{H}{I} masses lower than this limit in UDGs in the Hydra I cluster of galaxies. Nevertheless, the impact of an undetected \ion{H}{I} mass on our dynamical modeling would be negligible, since according to the available extension of the kinematic data, the traced radial range ($\sim0.9-1.5\,R_{\rm eff}$) is dominated by the stellar component, as any \ion{H}{I} gas in these galaxies is typically distributed well beyond these radial extensions.

\subsection{Asymmetric drift and thickness correction}
\label{sec:asymmetric_drift}
UDG-1 and LSB-6 are indeed systems supported by a combination of both ordered rotation and random motions. A more complex and rigorous description of the internal dynamics is thus required to address both the underlying kinematics and three-dimensional geometry and to recover the true circular velocities of stars. First, it is necessary to correct the stellar streaming motion for asymmetric drift \citep[AD, ][]{Binney2008} to properly account for the pressure support provided by the velocity dispersion. Second, the dynamical model should incorporate a realistic physical thickness for the stellar distribution. Neglecting the thickness of the galaxy would impact the inferred results in the modeling, yielding overestimated values for circular velocity \citep{Iorio2017}. As reported by \cite{Mancera-Pina2025}, the impact of the thickness correction becomes important in systems with $M_\ast\lesssim10^7M_\odot$ with $v_{\rm rot}/\sigma_{\rm HI}\lesssim3$. Similarly, neglecting the AD correction term would result in an underestimation of the global circular velocity of the stars and subsequently of the enclosed dynamical mass; this correction becomes negligible only in systems where $v/\sigma\gg1$ \citep{Iorio2017, Mancera-Pina2025}. Following these considerations, the measured line-of-sight velocity is defined as
\begin{equation}\label{eq:vlos}
    v_{\mathrm{los}}(R) = v_{\rm rot}(R) \sin i \cos \varphi \, ,
\end{equation}
where $v_{\rm rot}$ is the stellar rotation velocity, $i$ is the inclination angle between the galaxy plane and the line-of-sight, and $\varphi$ is the position angle of the photometric major axis, measured counterclockwise from north. The projected radius $R$ in the plane of the sky is defined through the coordinate transformation \citep{Lelli:2023wjh}
\begin{equation}
    \begin{aligned}
        x' &= (x \cos\varphi + y \sin\varphi) \cos i \, ,\\
        y' &= - x \sin\varphi + y \cos\varphi \, ,\\
        z' &= (x \cos\varphi + y \sin\varphi) \sin i \, ,\\
        R &= \sqrt{(x')^2 + (y')^2 + (z')^2} \, ,
    \end{aligned}
\end{equation}
with $\cos\varphi = y'/R$. Here, $(x,y,z)$ denote the coordinates in the galaxy reference frame, while $(x',y',z')$ correspond to the observer’s frame, where the $x'y'$ plane is orthogonal to the line of sight. The rotation velocity is defined as
\begin{equation}
    v^2_{\rm rot} = v_c^2 - v^2_{\rm AD} \, ,
\end{equation}
where $v_c$ is the circular velocity of the stars on the galactic plane and $v_{\rm AD}$ is the AD correction term \citep{Binney2008}. They are respectively defined as
\begin{equation}\label{eq:circ_vel_AD}
    v_c^2(R) = R \left( \frac{\partial \Phi}{\partial R} \right) \Bigg|_{z=0} \, , \qquad v^2_{\rm AD}(R) = -\frac{R}{\rho} \frac{\partial(\rho \sigma_R^2)} {\partial R} \, ,
\end{equation}
with $\Phi$ the gravitational potential of the galaxy evaluated at $z=0$, $\rho$ the volumetric mass density of the tracer, i.e., the stars, and $\sigma_R$ the corresponding radial velocity dispersion. Given the limitations in the spatial resolution of our kinematic data due to the extremely faint nature of UDGs, we adopted a simplified, yet physically motivated framework to evaluate the AD correction term. For both galaxies, we assumed that the stellar component has a thick, non-flaring distribution with constant intrinsic flattening ($q_0$). This assumption allows us to express the volumetric mass density of the stars $\rho$ in terms of the stellar surface intensity $I$ through the relation
\begin{equation}
    \rho \propto \frac{I \,\cos i}{\Upsilon_* \,h_z} \, ,
\end{equation}
where $\cos^2 i = (q^2 - q_0^2)/(1 - q_0^2)$, $\Upsilon_*$ is the stellar mass-to-light ratio and $h_z$ is the vertical scale length. Moreover, we assume velocity isotropy ($\sigma_R=\sigma_z=\sigma_\phi$) to remove any radial dependence of velocity dispersion ($\partial \ln \sigma_R^2 / \partial R = 0$). We verified the impact of this by assuming a radial anisotropy of $\beta = 0.2$ (typical of dwarf ellipticals, \citealp{Lipka2024}) instead of isotropy. The resulting circular velocity profiles are consistent within uncertainties for both galaxies, confirming that our isotropic assumption does not bias the derived rotation velocities.
The AD term is simplified as
\begin{equation}
       v^2_{\rm AD}(R) \simeq  -R\,\sigma^2 \,\frac{\partial\,{\rm ln}(I)}{\partial R} \, .
\end{equation}

The stellar surface brightness distribution is described using a Sérsic profile \citep{1968adga.book.S, Graham:2005fy}
\begin{equation}\label{Sersic}
    I(R) = I_{\rm eff}  \exp \bigg \{ -b_n \bigg[ \bigg( \frac{R}{R_{\rm eff}} \bigg)^{1/n} -1\bigg] \bigg\} \, ,
\end{equation}
where $I_{\rm eff}$ is the effective intensity, i.e., the surface brightness measured at effective radius $R_{\rm eff}$, the Sérsic index $n$, which determines the overall shape of the profile and the slope of the outer regions, and $b_n$, a dimensionless parameter defined as a function of $n$ to ensure that $R_{\rm eff}$ encloses exactly half the total light. The surface brightness can be additionally expressed in terms of $\log_{10}I(R) \propto -0.4\mu(R)$, and can be directly derived from photometric analysis of fitted isophotes of the galaxies.

We assumed an intrinsic thickness of $q_0=0.4$, a typical flattening value for dwarf-like systems with stellar masses $M_* < 10^9\,M_\odot$ \citep{Benavides2026}, while we adopted ($n$, $R_{\rm eff}$, $\mu_{\rm eff}$) from \cite{Iodice_2020} and \cite{LaMarca2022b} to compute the radial gradient of the stellar distribution. The values of velocity dispersion have been computed from the 2D Voronoi binning map derived in \cite{Buttitta_2025}. The stellar velocity map $v_{\rm los}$ derived in this work has a more extended kinematic information coverage; thus, there is no exact bin correspondence between the Voronoi tessellation map of $v_{\rm los}$ and that of $\sigma_{\rm los}$ previously derived in \cite{Buttitta_2025}. To assign a velocity dispersion value to each bin of the $v_{\rm los}$ map, we computed $\sigma_{{\rm bin},v}$ as the weighted mean of the values from all $\sigma_{i}$ bins spatially overlapping the given $v_{\rm los}$ bin weighted by the number of overlapping pixels ($N_i$) contributed by each $\sigma_i$ bin
\begin{equation}
    \sigma_{{\rm bin},v} = \frac{\sum_i \sigma_i N_i}{\sum_i N_i} \, .
\end{equation}
The uncertainties on $v_{\rm AD}$ have been computed by using standard error propagation formulas accounting for independent errors on $v_{\rm los}$ and $\sigma_{\rm los}$.

To complete the analysis, we additionally computed the intrinsic kinematic ratios. The galaxy inclination has been derived as  
\begin{equation}
    i = \arccos\left(\sqrt{\frac{q^2 - q_0^2}{1 - q_0^2}}\right)
,\end{equation}
with $q$ the projected axial ratio derived from photometric analysis ($q = 1 -\epsilon$) and $q_0=0.4$ the intrinsic flattening as adopted in the modeling formalism. We obtained $i\sim44^\circ$ and  $i\sim69^\circ$ for UDG-1 and LSB-6, respectively. The intrinsic dynamical support has been derived by deprojecting the velocity values into the galaxy plane: $(v/\sigma)_{\rm in}=v_{\rm R_{max}}/(\sin i \cdot \sigma_{\rm eff})$. We thus recovered $(v/\sigma)_{\rm in}\sim3.4$ for UDG-1 and $(v/\sigma)_{\rm in}\sim1.7$ for LSB-6. The resulting value of the intrinsic kinematic ratio places UDG-1 in a regime where the effect of the thickness correction is less important, whereas for LSB-6 the thickness correction, if not accounted for, might introduce an important bias.

\subsection{Stellar component}\label{sec:star_modeling}

For the stellar components of UDG-1 and LSB-6, we approximated the mass distribution as a homogeneous oblate spheroid following a Maclaurin law \citep{Binney2008}. This choice is motivated by the fact that these extremely diffuse systems are typically described by Sérsic profiles with indices in the range $n \sim 0.5-1.5$ \citep{LaMarca2022b}, resulting in relatively flat and slowly declining light profiles that can, to first order, be approximated by a constant-density spheroid. Within this framework, the stellar contribution to the circular velocity is given by
\begin{equation}
\begin{aligned}
v_c^2(R) &= \Omega_0 R^2 \, , \\
\Omega_0 &= 2\pi G \rho_0 \left( \frac{\sqrt{1 - e^2}}{e^3} \arcsin(e) - \frac{1 - e^2}{e^2} \right) \, ,\\
\rho_0 &= \frac{3\,\Upsilon_* \,L_{\text{tot}}}{4\pi \cdot a^3 \cdot q_0} \, , \\
e &= \sqrt{1 - q_0^2} \, ,
\end{aligned}
\end{equation}
where $L_{tot}$ is the total luminosity, $e$ is the eccentricity, and $a$ is the equatorial semi-axis of the spheroid. We set the intrinsic flattening equal to $q_0=0.4$, consistent with the assumptions adopted for the AD correction (Sec.~\ref{sec:asymmetric_drift}), while the semi-axes of the two galaxies have been retrieved from photometric analysis of fitted isophotes of the galaxy \citep{Iodice_2020, LaMarca2022b}. This formalism is valid under the condition $R\leq a$. In our case, this condition is always valid since the kinematic spatial coverage is less extended than the axis $a$ of the galaxy: for UDG-1 and LSB-6, the extension of kinematic data reaches $R_{\rm max}\sim1.5R_{\rm eff}\sim2.6$ kpc and $R_{\rm max}\sim0.9R_{\rm eff}\sim3.6$ kpc, respectively; while the photometric major axes are $a=8$ kpc and $a=10$ kpc, respectively.

\subsection{Dark matter component}\label{sec:DM_modeling}
The dominant contribution to the galaxy kinematics arises from the gravitational potential generated by the DM halo. Regardless of the specific DM particle candidate, assuming spherical symmetry provides a good approximation for the halo structure. Under this assumption, the circular velocity defined in Eq.~\ref{eq:circ_vel_AD} can be written as
\begin{equation}\label{eq:vc_dm}
    \begin{aligned}
        v_c^2(R) = &-2\pi G R \, \frac{\partial}{\partial R} \Bigg( \int_0^R dr \, r^2  \rho_{\mathrm{dm}}(r) \\
        & \times\int_0^\pi d\theta \, \frac{\sin\theta}{\sqrt{R^2 + r^2 -  rR\cos\theta}} + \int_R^\infty dr \, r^2 \rho_{\mathrm{dm}}(r) \\
        &\times \int_0^\pi d\theta \, \frac{\sin\theta}    {\sqrt{R^2 + r^2 - rR\cos\theta}} \Bigg)\Bigg|_{z=0} \, ,
    \end{aligned}
\end{equation}
where $\rho_{\mathrm{dm}}(r)$ denotes the DM density profiles introduced in the previous section. For a spherically symmetric mass distribution, however, Newton’s shell theorem allows Eq.~\ref{eq:vc_dm} to be expressed in the simpler form
\begin{equation}
    v_c^2(R) = \frac{GM(R)}{R} = \frac{4 \pi G}{R} \int_0^R dr \, r^2 \rho_{\mathrm{dm}}(r) \, ,
\end{equation}
where $M(R)$ is the mass enclosed within radius $R$.

\section{Statistical analysis}\label{sec:statistical_analysis}

The aim of this work is to constrain the parameters that describe the internal structure of UDG-1 and LSB-6. This allows us to investigate which DM model and UDG formation scenario are preferred by the observed kinematic data.

\subsection{Likelihood and priors}\label{sec:priors}

We performed a Bayesian analysis based on a likelihood comparison at the level of two-dimensional stellar velocity fields. In this framework, both the observed and theoretical velocities are functions of the position $(x,y)$ on the velocity maps, rather than of the galactocentric distance $R$. We adopted a Gaussian likelihood, with the corresponding $\chi^2$ function given by
\begin{equation}
    \chi^2(\mathbf{\theta}) = \sum_{i=1}^{\mathcal{N}_{\mathrm{data}}} \left[ \frac{v_{\mathrm{los},i}(\mathbf{\theta}) - v_{\mathrm{obs},i}}{\delta_{\mathrm{obs},i}} \right]^2 \, .
\end{equation}
Here, $v_{\mathrm{obs},i}$ and $\delta_{\mathrm{obs},i}$ denote the observed line-of-sight velocity and its associated uncertainty, respectively, both including the AD correction (Sec.~\ref{sec:asymmetric_drift}), while $v_{\mathrm{los},i}(\mathbf{\theta})$ is the model prediction for the circular velocity projected along the line of sight, computed according to Eq.~\ref{eq:vlos}. The vector $\mathbf{\theta}=\{ \text{DM parameters}, \Upsilon_*\}$ represents the set of free parameters sampled in the statistical analysis. The specific DM parameters included in $\mathbf{\theta}$ depend on the model under consideration and follow the definitions given in Sec.~\ref{sec:dark_matter}.

To ensure a robust fit and mitigate potential biases arising from outliers, we applied a data-selection procedure before the Markov Chain Monte Carlo (MCMC) analysis. Specifically, during the likelihood evaluation, we retained only those data points whose observed velocities $v_{\mathrm{obs},i}$ are consistent with an ordered rotational motion of the galaxy within $8 \,\delta_{\mathrm{obs},i}$. This conservative threshold rejects $\lesssim$ 18\% of the data points for both galaxies, ensuring that only extreme outliers -- likely arising from localized noncircular motions or observational artifacts -- are excluded.

Throughout the analysis, we adopted prior distributions that encode additional information on the sampled parameters. We impose a Gaussian prior on the stellar mass-to-light ratio, $\Upsilon_*$, based on recent stellar population modeling of the LEWIS galaxies \citep{Doll2026} and derived from $r$-band synthetic photometry from stellar template spectra. Specifically, we assumed 
\begin{equation}
    \Upsilon_* = 
    \begin{cases}
        \mathcal{N}(1.5, \, 1.2) & \text{for UDG-1} \, ,\\
        \mathcal{N}(3.1, \, 0.6) & \text{for LSB-6} \, ,
    \end{cases}
\end{equation}
using a dispersion three times larger than that inferred by \cite{Doll2026}. This prior on $\Upsilon_*$ is applied in all cases except when the galaxies are modeled as DM deficient systems, i.e., containing only a stellar component.

Moreover, whenever we included a DM component -- either standard or non-standard -- we also imposed a Gaussian prior based on the halo mass-concentration relation. When a standard CDM component is considered, we included prior information from the $c$-$M$ relation derived by \cite{Correa:2015dva} within the $\Lambda$CDM cosmology. This relation, based on a semi-analytic model that combines an analytic description of halo mass growth with an empirical concentration-formation time relation calibrated on N-body simulations, is given by
\begin{equation}
    \log_{10}c_{200} = \alpha + \beta \log_{10}(M_{200}) \left[ 1 + \gamma \log_{10}(M_{200})^2 \right] \, ,
\end{equation}
where
\begin{equation}
    \begin{aligned}
        \alpha &= 1.7543 - 0.2766 \, (1 + z) + 0.02039 \, (1 + z)^2 \, , \\
        \beta &= 0.2753 + 0.00351 \, (1 + z) - 0.3038 \, (1 + z)^{0.0269} \, , \\
        \gamma &= -0.01537 + 0.02102 \, (1 + z)^{-0.1475} \, ,
    \end{aligned}
\end{equation}
and $z$ is the redshift of the galaxy. The prior is modeled as a Gaussian distribution in log space, with a dispersion of $0.16$ dex.

When a FDM component is considered, we included prior information from the $c$-$M$ relation of \cite{Marsh:2016vgj}, which accounts for the characteristic suppression of structure formation in models with truncated linear power spectra. The resulting modified mass-concentration relation is
\begin{equation}
    c_{200}^{\mathrm{FDM}} \left( M_{200}, m_{\alpha} \right) = c_{200}^{\mathrm{CDM}} \left( M_{200} \right) \left[ 1 + \gamma_1 \frac{M_{1/2} (m_{\alpha})}{M_{200}} \right]^{-\gamma_2} \, ,
\end{equation}
where we adopted the standard CDM $c$-$M$ relation from \cite{Correa:2015dva}, set $\gamma_1 = 15$ and $\gamma_2 = 0.3$, and defined
\begin{equation}
    \begin{aligned}
        M_{1/2} (m_{\alpha}) &= \frac{4}{3} \pi \, \Omega_m \, \rho_{\mathrm{c}} \left[ \frac{\pi}{k_{1/2}(m_{\alpha})} \right]^{3} \, , \\
        k_{1/2}(m_{\alpha}) &= 4.5 \left( \frac{m_\alpha}{10^{-22} \mathrm{eV}} \right)^{4/9} \mathrm{Mpc}^{-1} \, .
    \end{aligned}
\end{equation}
The prior is modeled as a Gaussian distribution in log space, with the same dispersion adopted in the standard CDM case.

When the galaxy model includes a SIDM component, we adopted the scaling relation proposed by \cite{Lin_2016}, which links the core properties to the halo mass within an analytic framework. This relation is almost independent of the underlying SIDM particle physics and is given by
\begin{equation}
    \rho_c r_c = \rho_s r_s = 41 \, M_\odot \, \mathrm{pc}^{-2} \left( \frac{M_{200}}{10^{10} M_\odot} \right)^{0.18} \, ,
\end{equation}
where the subscript $c$ denotes core quantities, while $s$ refers to the corresponding NFW parameters. The relation is implemented as a Gaussian prior in logarithmic space, adopting the same dispersion used for the standard CDM and FDM halo mass-concentration relations.

Finally, when a NMC between DM and gravity is considered, we adopted the same priors used in the standard CDM analysis. This choice is motivated by the fact that the NMC scenario is generally intended to preserve the successful $\Lambda$CDM phenomenology on large scales, while allowing deviations to appear mainly in the smaller, nonlinear scales, where the dynamical constraints are applied \citep{Bettoni:2012xv}. A comprehensive treatment of structure formation in the NMC framework would require the inclusion of nonlinear perturbations and dedicated N-body simulations, which are beyond the scope of this work. We note, however, that this simplifying assumption may bias the analysis toward minimal deviations from $\Lambda$CDM.

\subsection{Sampling and model comparison}\label{sec:bayes}

For the MCMC analysis, we used the affine-invariant ensemble sampler \citep{ensamble_sampler} in the $\mathtt{emcee}$ package \citep{emcee}. Convergence of the chains is assessed through the integrated autocorrelation time \citep{Sokal1997}. 

To compare the different DM models and quantify their ability to reproduce the observed kinematics of UDG-1 and LSB-6, we computed the Bayes factor $\mathcal{B}_{ij}$, where $\mathcal{M}_j$ denotes the reference model against which model $\mathcal{M}_i$ is tested. The Bayes factor is defined as the ratio of the Bayesian evidence of the two models. The evidence is given by
\begin{equation}
    \mathcal{Z} = \int \mathcal{L}(\theta) \, \pi (\theta) \, \mathrm{d}\theta \, ,
\end{equation}
where $\mathcal{L}$ is the likelihood, $\pi$ the prior, and $\theta$ the parameter vector. We evaluated the Bayesian evidence using the nested sampling algorithm introduced by \cite{Mukherjee:2005wg}. Because the algorithm is stochastic, we ran it 100 times to reduce statistical noise. This procedure yields a distribution of evidence values, from which we report in Table~\ref{tab:UDG1} and \ref{tab:LSB6} the median Bayes factor and its associated uncertainty. Model selection is then performed according to the Jeffreys' scale \citep{Jeffreys:1939xee}:
\begin{equation*}
    \begin{aligned}
        \ln \mathcal{B}_{ij} < 0  \qquad & \text{evidence against model $\mathcal{M}_i$} \, , \\
        0 < \ln \mathcal{B}_{ij} < 1 \qquad & \text{inconclusive evidence} \, , \\
        1 < \ln \mathcal{B}_{ij} < 2.5 \qquad & \text{substantial evidence in favor of $\mathcal{M}_i$} \, , \\
        2.5 < \ln \mathcal{B}_{ij} < 5 \qquad & \text{strong evidence in favor of $\mathcal{M}_i$} \, , \\
        \ln \mathcal{B}_{ij} > 5 \qquad & \text{decisive evidence in favor of $\mathcal{M}_i$} \, .
    \end{aligned}
\end{equation*}

Model comparison based on Bayesian evidence requires particular care when the overall statistical quality of the fits is poor. In such cases, residual outliers can drive the reduced $\chi^2$ to relatively large values, so that even small changes in the fit may produce substantial variations in the absolute $\chi^2$. Because the Bayesian evidence depends exponentially on the $\chi^2$, these variations can translate into large differences in the Bayes factor, despite representing only small relative changes in the goodness-of-fit. Consequently, the interpretation of Bayes factors, particularly in the context of the Jeffreys' scale, should be approached with caution to avoid overestimating the significance of the results.

\section{Results}\label{sec:results}

The results of the statistical analysis for the two LEWIS galaxies are presented in Table~\ref{tab:UDG1} for UDG-1 and Table~\ref{tab:LSB6} for LSB-6. For each model, we report the inferred values (median $\pm 1 \sigma$) of the parameters describing the internal structure of the galaxies, together with the values of $\chi^2_L$ associated with the likelihood and $\chi^2_P$ related to the posterior, both evaluated at the median values of the inferred parameters, and the Bayes factor $\ln\mathcal{B}$. To compute the latter, the reference model is the standard CDM scenario, described by the soft-cored Einasto profile defined in Eq.~\ref{Einasto}. Since the qualitative behavior of the predicted velocity fields is common to all DM models, both standard and non-standard, and the resulting two-dimensional velocity fields are nearly indistinguishable across models, we show only the best-fitting Einasto maps (Fig.~\ref{fig:2Dvelocitymap}) to avoid redundancy. The corresponding residual maps for all tested models are shown in Appendix~\ref{AppendixC:residuals}.

In the following sections, we describe the constraints on the DM candidates and the halo profiles.

\subsection{Dark matter free scenario}\label{subsec:DM_free_results}

Ultra-diffuse galaxies exhibit a wide range of DM contents, from strongly DM-dominated systems to galaxies that appear to lack DM \citep{vanDokkum2018, vanDokkum2019, Gannon2021, Buzzo2024, Gannon2024, Buttitta_2025}. Adopting an agnostic approach, we first examine the possibility that UDG-1 and LSB-6 are DM-free systems, modeling them solely through their stellar components. The first sections of Table~\ref{tab:UDG1} and Table~\ref{tab:LSB6} report the corresponding constraints and goodness-of-fit for this scenario.

The fit to the UDG-1 data is statistically equivalent to the reference model that includes a DM halo, whereas the LSB-6 data provide strong evidence against a model without any DM component. In both cases, however, reproducing the observed kinematics without DM requires stellar mass-to-light ratios roughly two orders of magnitude above those derived from stellar population modeling in \cite{Doll2026}, corresponding to $7\sigma$ and $19\sigma$ discrepancies for UDG-1 and LSB-6, respectively. Such values are also inconsistent with the $\Upsilon_* \sim 0.5-3.5\, M_\odot/L_\odot$ range typically obtained from full spectral fitting and Bayesian SED analyses of UDGs in cluster environments (e.g., Virgo, Hydra, Coma, Perseus; \citealp{Ferre-Mateu2018, Ferre-Mateu2023, Buzzo2024, Gannon2024, Hartke2025}). We thus conclude that a DM-free scenario can be robustly excluded for both LEWIS galaxies.

\subsection{Standard cold dark matter}\label{sec:standardCDM_results}

\begin{figure*}
    \centering
    \includegraphics[scale=0.79]{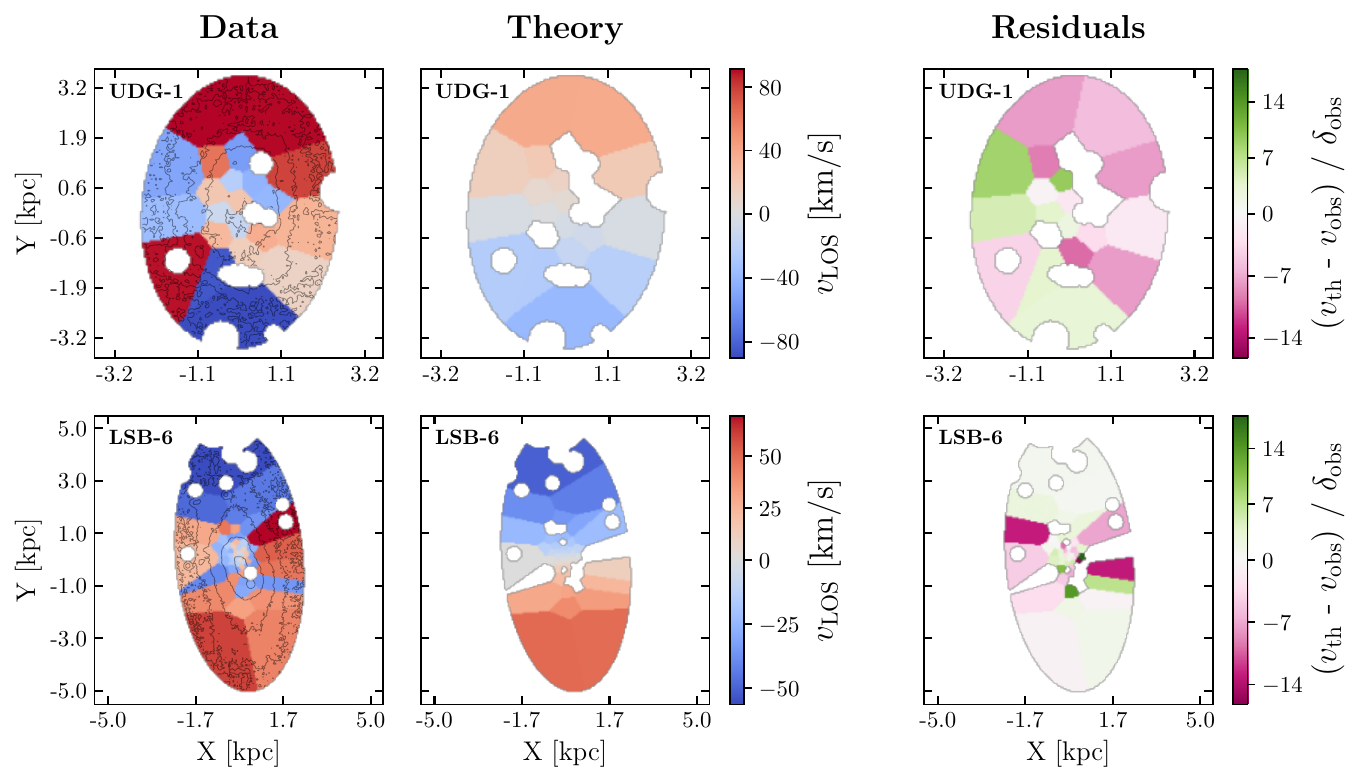}
    \caption{Velocity fields for UDG-1 (\textit{top}) and LSB-6 (\textit{bottom}). \textit{Left panels}: Observed line-of-sight velocities overlaid on flux contours from the white image extracted from the MUSE datacube of the galaxy. \textit{Central panels}: Velocity maps predicted by the best-fitting model for the standard dark matter framework, the Einasto profile. \textit{Right panels}: Corresponding normalized residual maps. All theoretical quantities are computed using the median parameter values reported in Tables~\ref{tab:UDG1} and~\ref{tab:LSB6}.}\label{fig:2Dvelocitymap}
\end{figure*}

We next consider a standard CDM modeling. We test a cuspy NFW profile (Eq.~\ref{NFW}), a cored Burkert profile (Eq.~\ref{Burkert}), and a soft-cored Einasto profile (Eq.~\ref{Einasto}).

The Einasto profile yields the best fit among the standard CDM models for both LEWIS galaxies. Figure~\ref{fig:2Dvelocitymap} compares the observed two-dimensional velocity fields of UDG-1 and LSB-6 with those predicted by the best-fitting Einasto models (Table~\ref{tab:UDG1} and \ref{tab:LSB6}), along with the corresponding normalized residual maps. The model reproduces the LSB-6 kinematics remarkably well, although a few outliers survived the data-selection procedure inflating the reduced $\chi^2$. For UDG-1, by contrast, the velocity field is systematically underestimated. This offset, which is consistently recovered for all the DM models explored in this work, may reflect limitations in the available kinematic data, such as sparse spatial sampling, residual noncircular motions, or limited radial coverage.

The soft-cored Einasto model yields a DM halo structure effectively equivalent to that inferred with the cored Burkert profile, which is only mildly disfavored by the data. The increase in $\chi^2_{P}$ and the negative Bayes factor is primarily driven by the smaller relative uncertainties on the Burkert DM halo parameters, which amplify the discrepancy with the adopted $c$-$M$ prior. Although the halo mass and concentration parameters reported in Table~\ref{tab:UDG1} and Table~\ref{tab:LSB6} differ between the two models, these discrepancies stem from the different definitions of halo concentration \citep{Kong:2022oyk} and from the extrapolation of the density profiles beyond the radial extent of the kinematic data. This is illustrated in Fig.~\ref{fig:density_profiles}, which compares the three-dimensional matter density profiles predicted by all DM models considered in this work: for both galaxies, the Einasto and Burkert profiles overlap over nearly the entire radial range probed by the observations. Distinguishing between the two parameterizations will therefore require kinematic measurements extending to significantly larger radii.

Moving on to the cuspy NFW profile, the halo model typically favored by CDM-only simulations, we find that it provides a statistically disfavored fit to the data. The strongly negative Bayes factors are primarily driven by the sensitivity of the predicted velocity field to the residual outliers: as shown in Fig.~\ref{fig:density_profiles}, the central cusp develops within the innermost $\sim1$ kpc, increasing the predicted circular velocities in the central regions of the stellar disk. Since this region contains most of the data points that deviate from ordered motion, the DM cusp amplifies the mismatch between model and observations, degrading the overall fit.

For both galaxies, the NFW concentration parameter $c_{200}$ tends toward relatively low values, consistent with previous studies of gas-rich UDGs \citep{ManceraPina:2019zih, PinaMancera:2021wpc, Sengupta:2019smr, Shi:2021tyg, Kong:2022oyk}. However, these concentrations are in disagreement with expectations for halos of similar mass, deviating by approximately $4\sigma$ from the $c$-$M$ relation of \cite{Correa:2015dva} for both UDG-1 and LSB-6.

\begin{figure*}
    \centering
    \includegraphics[scale=0.5]{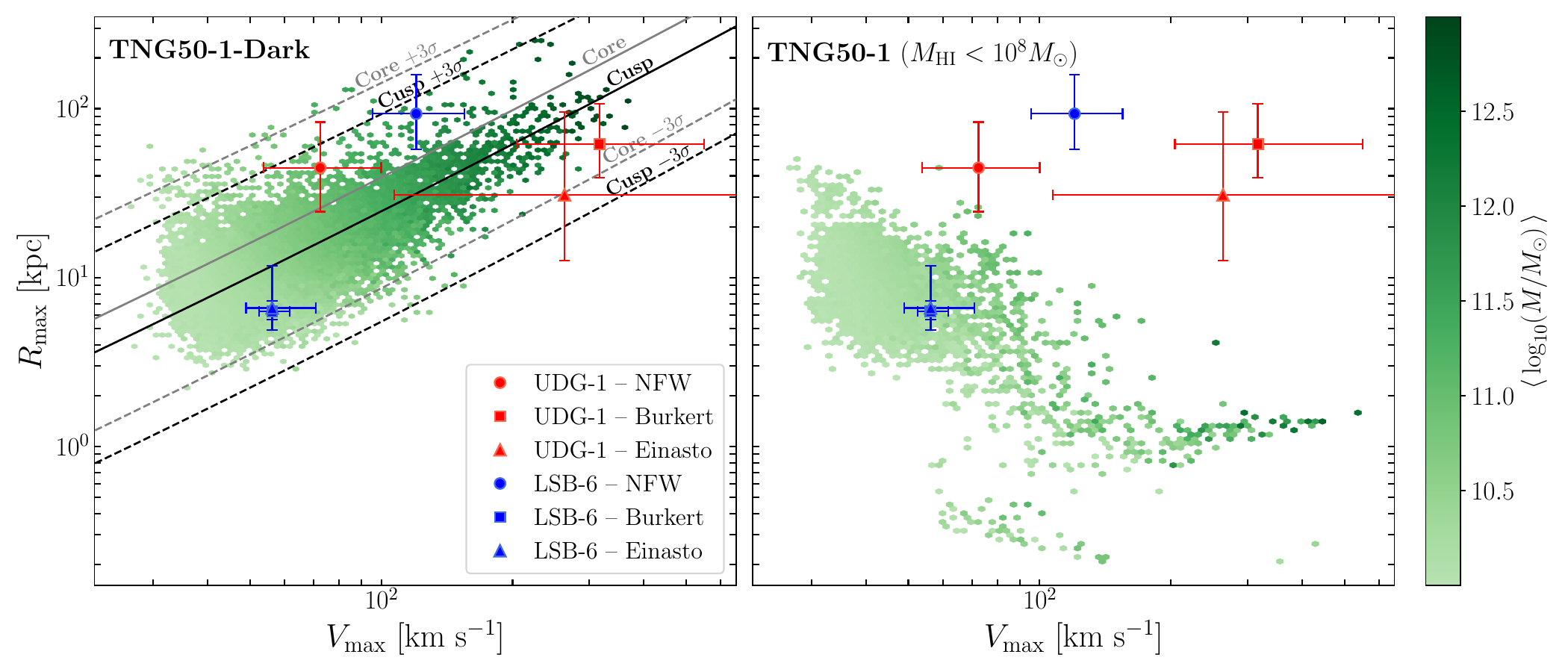}
    \caption{Distribution of simulated dark matter halos in the $R_{\max}$-$V_{\max}$ plane from the IllustrisTNG \citep{Nelson:2018uso} dark-matter-only (\textit{left}) and hydrodynamical (\textit{right}) simulations, color-coded by halo mass. The observational constraints are overlaid for the LEWIS galaxies -- UDG-1 (red) and LSB-6 (blue) -- derived from dynamical modeling assuming NFW (circles), Burkert (squares), and Einasto (triangles) profiles, respectively. In the left panel, the solid and dashed curves show the median and median $\pm3\sigma$ $R_{\max}$-$V_{\max}$ relations for cuspy NFW (black) and cored Burkert (gray) halos, as derived from \cite{Correa:2015dva}.}\label{fig:TNG}
\end{figure*}

To further assess the validity of the standard CDM framework for the two LEWIS galaxies, in Fig.~\ref{fig:TNG} we compare the inferred halo properties with those of simulated halos from the IllustrisTNG project \citep{Nelson:2018uso}. Following \cite{Kong:2022oyk}, we use the $z=0$ snapshot (snapshot 99) of the publicly available TNG50-1-Dark (dark-matter-only) and TNG50-1 (hydrodynamical) simulations, selecting halos with $\log_{10}(M_{200}/M_{\odot}) \in (10.0, 13.0)$. For the hydrodynamical sample, we additionally impose a \ion{H}{I} mass cut ($<10^8\,M_\odot$) to approximately match the gas content of the galaxies analyzed in this work (Sec.~\ref{sec:galaxy_modeling}). For each halo, we retrieve the catalog values of the maximum circular velocity, $V_{\mathrm{max}}$, and the corresponding radius, $R_{\mathrm{max}}$. In the left panel of Fig.~\ref{fig:TNG}, we also show the theoretical $V_{\mathrm{max}}-R_{\mathrm{max}}$ relations that we derived from \cite{Correa:2015dva} for cuspy and cored halo profiles.

Figure~\ref{fig:TNG} shows that both the Einasto and Burkert models inferred for UDG-1 occupy a region of the $V_{\mathrm{max}}-R_{\mathrm{max}}$ plane not populated by simulated halos, and thus physically implausible within the CDM framework. This discrepancy is particularly evident in the hydrodynamical simulation, where baryonic effects are explicitly included and the Einasto and Burkert profiles lie at $\sim 2 \sigma$ and $\sim 3 \sigma$, respectively, from the populated region. This tension originates from the extremely massive cored halo required to reproduce the observed kinematics of UDG-1. In the Burkert case, we infer $M_{200}=2.7^{+11.1}_{-2.0} \times 10^{13} M_{\odot}$, corresponding to a star-to-halo mass ratio of $\log_{10}(M_{*}/M_{200})=-5.4^{+0.6}_{-0.7}$, which is in strong disagreement with the empirical scaling relations for LSB galaxies of \cite{Di_Paolo_2019} given the stellar mass of UDG-1, $M_{*}=1.12 \times 10^{8} M_{\odot}$ \citep{Iodice_2023}. Moreover, the inferred core radius, $r_s = 18.60^{+13.00}_{-6.94}$ kpc, extends well beyond the radial extent of the kinematic data, implying a severe extrapolation. A similar tension arises within the Einasto framework, where the shape parameter is driven to an extremely low value, $n_{\mathrm{E}}=1/\gamma < 0.17$, in strong disagreement with the constraints reported by \citet{Di_Cintio_2013} for dwarf galaxies.

In contrast, the NFW profile provides the most physically plausible description of the DM halo, despite yielding a poorer goodness-of-fit. As shown in Fig.~\ref{fig:TNG}, the NFW solution occupies a region of the $V_{\mathrm{max}}-R_{\mathrm{max}}$ plane that is largely populated in the dark-matter-only simulation and still partially populated in the hydrodynamical run. Thus, although the NFW halo required by UDG-1 has relatively rare and extreme properties, it remains compatible with the range of halo structures produced in cosmological simulations.

On the other hand, for LSB-6, the halo properties inferred from both cuspless profiles are fully consistent with the distribution of simulated halos (Fig.~\ref{fig:TNG}), so that both parameterizations provide a suitable description of its DM halo. For the Einasto model, the shape parameter, $n_{\mathrm{E}}=1.14^{+0.56}_{-0.33}$, is consistent at the $1\sigma$ level with the constraints of \citet{Di_Cintio_2013} for dwarf galaxies. In the Burkert case, the inferred core radius, $r_s = 1.93^{+0.27}_{-0.22}$ kpc, lies well within the radial range probed by the observations and is thus robustly constrained. The resulting star-to-halo mass ratio, $\log_{10}(M_{*}/M_{200})=-2.7^{+0.1}_{-0.1}$, places LSB-6 close to the intersection between the $\log_{10}(M_{*}/M_{200})-M_{*}$ scaling relations of LSB and dwarf galaxies (see Fig.~11 of \citealp{Di_Paolo_2019}). Conversely, the kinematic properties of the inferred NFW halo are in strong disagreement with the halo population found in the TNG50-1 hydrodynamical simulation. These results support the interpretation that LSB-6 has a cored DM halo, consistent with the well-established cusp-core tension observed in LSB galaxies.

It is also worth noting that, irrespective of the adopted standard CDM profile, the dynamical mass inferred for UDG-1 is slightly lower than the value reported by \cite{Buttitta_2025}, although the two estimates remain consistent within $1\sigma$. For LSB-6, instead, the inferred dynamical mass is in excellent agreement with that derived by \cite{Buttitta_2025}. Moreover, the reference standard CDM models adopted for the two LEWIS galaxies -- namely, a cuspy profile for UDG-1 and a cuspless profile for LSB-6 -- are also consistent with the stellar-to-halo mass relation (SHMR) derived by \cite{SHMR_relation}. This relation connects the stellar mass of a galaxy to the mass of its host halo and is derived from a semi-empirical model describing the co-evolution of galaxies and halos over a wide range of redshifts and masses within the $\Lambda$CDM framework. For our systems, the inferred stellar-to-halo mass ratios are consistent with the expected SHMR within approximately $2\sigma$ for UDG-1 and $1\sigma$ for LSB-6, respectively. These deviations are fully compatible with observational estimates of the scatter at the low-mass end of the SHMR \citep{2023ApJ...956....6D, ManceraPina:2025szx}.

\subsection{Fuzzy dark matter}

\begin{figure}[!t]
    \centering
    \includegraphics[scale=0.3]{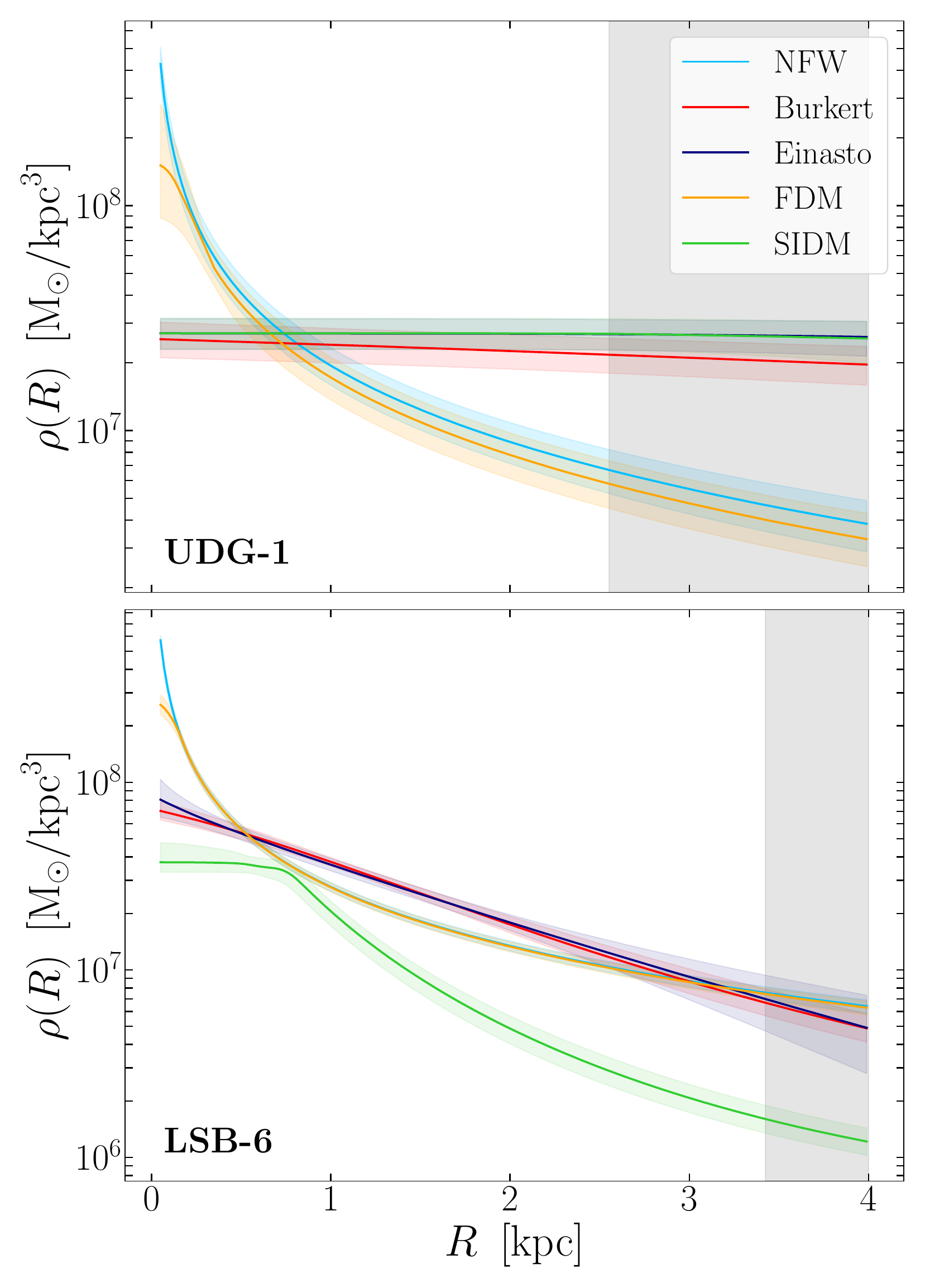}
    \caption{Three-dimensional matter density profiles for UDG-1 (\textit{top}) and LSB-6 (\textit{bottom}). Solid lines show median profiles for each model, with shaded regions indicating $1\sigma$ uncertainties. The gray shaded areas mark radial ranges not covered by the observational data.}\label{fig:density_profiles}
\end{figure}

Among the non-standard DM scenarios, FDM emerges as a viable alternative. It reproduces the observed velocity fields with a statistical quality comparable to that of the NFW model. Consequently, FDM is statistically equivalent to the reference model for UDG-1, namely the cuspy NFW profile, but disfavored with respect to the cuspless reference models for LSB-6. As shown in Fig.~\ref{fig:density_profiles}, the FDM density profiles closely resemble the corresponding NFW profiles over the radial range probed by the data, owing to the small size of their solitonic cores. As a result, the inferred halo mass, concentration, and dynamical mass are all consistent with the NFW framework. Unlike in the standard CDM interpretation, however, the relatively low halo concentrations inferred in the FDM scenario are fully consistent with the expected $c$-$M$ correlation for FDM systems \citep{Marsh:2016vgj}.

Regarding the ALP mass, the constraints from the two LEWIS galaxies are mutually consistent, yielding $m_\alpha = 4.74^{+1.83}_{-1.14} \times 10^{-22}$ eV for UDG-1 and $m_\alpha = 6.26^{+0.41}_{-0.38} \times 10^{-22}$ eV for LSB-6. These values are fully consistent with those inferred from the stellar kinematics of the DM-dominated UDG Dragonfly~44 \citep{Wasserman_2019}, as well as with the constraints derived from the half-light masses of the dSph galaxies Draco II and Triangulum II \citep{Calabrese:2016hmp}. Likewise, the ALP mass inferred by \cite{Chen:2016unw} from a sample of eight dSph galaxies is consistent with our UDG-1 estimate within $2.5\sigma$, though in $\sim10\sigma$ disagreement with our LSB-6 value. However, our measurements are in significant tension with several other determinations reported in the literature. The ALP masses inferred for the DM-dominated UDG Nube \citep{Montes:2023ahn} and for AGC~114905 \citep{Mancera-Pina2024} -- a UDG with an exceptionally high baryon fraction -- are approximately one order of magnitude smaller than our estimates, as are the constraints obtained by \cite{Banares-Hernandez:2023axy} for thirteen nearby dwarf irregular galaxies and by \cite{Bernal_2017} for a large sample of LSB galaxies. In contrast, a recent study of a Milky Way dwarf satellite by \cite{Zimmermann:2024xvd} places a $2\sigma$ lower limit of $m_\alpha > 2.2 \times 10^{-21}$ eV, approximately one order of magnitude higher than our measurement.

Our results can also be compared with the scaling relations presented in \cite{Banares-Hernandez:2023axy}. The measurements for both UDG-1 and LSB-6 lie significantly above their empirical $m_\alpha$-$M_*$ relation, based on a relatively small sample of nearby dwarf irregular galaxies. Moreover, our galaxies do not follow the predicted SHMR \citep{Banares-Hernandez:2023axy}: for halo masses above $5 \times 10^{10} M_{\odot}$, the expected stellar masses are roughly $3\sigma$ larger than those we infer for the two LEWIS galaxies, which instead agree with the values reported by \cite{Iodice_2023} ($\sim 1.5 \times 10^8 M_\odot$). Collectively, these tensions suggest that current FDM models face difficulties in simultaneously reproducing the observed properties of LSB systems. That said, given the still limited number of galaxies tested against FDM predictions, as well as the rich theoretical FDM landscape yet to be fully explored, we consider the issue to remain open and worthy of further investigation.

Finally, our ALP mass constraints can be compared with independent results from cosmological probes. The least stringent bounds -- still compatible with most constraints derived from galaxy kinematics -- come from combined analyses of cosmic microwave background and weak lensing data, yielding a $2\,\sigma$ lower limit of $m_\alpha > 10^{-23}$ eV \citep{Dentler:2021zij}. Tighter constraints arise from studies of ultraviolet luminosity functions (UVLFs), which place a lower bound of $m_\alpha > 2.5 \times 10^{-22}$ eV at the $2\,\sigma$ level \citep{Winch:2024mrt}. This bound is fully consistent with our measurements and with those for Dragonfly~44 and most of the dSph galaxies discussed above. By contrast, it is in disagreement with the lower ALP masses reported for other UDGs, LSB galaxies, and dwarf irregular systems. Even stronger limits are inferred from analyses of the high-redshift turnover of the UVLF using gravitationally lensed galaxies, with \cite{Sipple:2024svt} reporting $m_\alpha > 1.5 \times 10^{-21}$ eV at $2\,\sigma$, thereby excluding the mass range favored by our analysis. Comparable bounds are obtained from Lyman~$\alpha$ forest studies, yielding a lower limit of $m_\alpha > 2.1 \times 10^{-21}$ eV at $2\,\sigma$ \citep{Nori:2018pka}. It is worth noting, however, that these cosmological analyses rely on multiple modeling assumptions that can significantly influence the resulting lower bounds.

\subsection{Self-interacting dark matter}

Another alternative is provided by the SIDM framework. In this scenario, the halo consists of a collisional core embedded within a collisionless outer halo. This parameterization successfully reproduces the kinematics of both LEWIS galaxies and yields the statistically preferred fit among all models considered in this work. Despite the good statistical agreement, however, the robustness of these results is affected by their dependence on the residual outliers present in the data.

\begin{figure}
    \centering
    \includegraphics[scale=0.8]{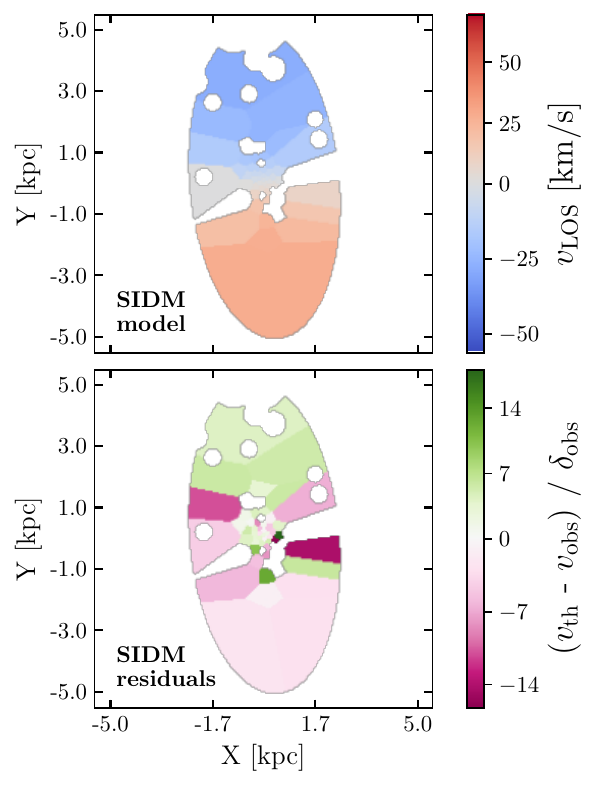}
    \caption{SIDM theoretical velocity field for LSB-6. \textit{Top}: Velocity maps predicted by the SIDM model using the median parameter values reported in Table~\ref{tab:LSB6}. \textit{Bottom}: Corresponding normalized residuals.} \label{fig:velocity_field_SIDM}
\end{figure}

For UDG-1, the SIDM model is statistically equivalent to the reference Einasto framework, with closely matching density profiles (Fig.~\ref{fig:density_profiles}). As in the standard cuspless models, however, the inferred halo properties are physically implausible: the collisional core extends well beyond the radial range probed by the kinematic data, making the inference strongly dependent on extrapolation. As a consequence of the broad core, the halo mass is extremely large, $M_{200}\simeq1.7 \times 10^{13} M_\odot$, implying that more than 99.999\% of the total UDG-1 mass would be in the form of dark matter. Using Eqs.~\ref{eq:SIDM_velocity} and \ref{eq:SIDM_cross_section}, we derive the characteristic relative velocity of the SIDM particles and the corresponding velocity-weighted self-interaction cross section
\begin{equation}
    \begin{aligned}
        \log_{10}v &= 2.43^{+0.33}_{-0.33} \,\,\,\, \mathrm{km} \, \mathrm{s}^{-1} \, , \\
        \frac{\langle \sigma v\rangle}{m} &= 22.78^{+4.26}_{-3.36} \,\,\,\, \mathrm{cm}^2 \, \mathrm{km} \, \mathrm{g}^{-1} \, \mathrm{s}^{-1} \, .
    \end{aligned}
\end{equation}
The inferred velocity-weighted cross section is broadly consistent with the values derived by \cite{Kaplinghat:2015aga} for dwarf and LSB galaxies. However, the corresponding relative particle velocity is more than a factor of two larger than expected from the velocity-dependent relation inferred from dwarf galaxies, LSB galaxies, and galaxy clusters. Rescaling the velocity-weighted quantity as $\langle \sigma v \rangle / m \, \times \, 1 / v_{\mathrm{median}}$ yields an approximate cross section $\sigma / m \approx 0.08 \, \mathrm{cm}^2 \, \mathrm{g}^{-1}$, which is in significant tension with the substantially larger values inferred by \cite{Almeida:2025cee} for ultra-faint dwarf galaxies. Together, these results indicate that the SIDM solution for UDG-1, while statistically viable, is not physically robust. However, this shortcoming stems from the limited quality and radial coverage of the available data, and the model cannot be completely excluded, thus warranting further investigation as more extended kinematic data become available.

For LSB-6, the SIDM model is statistically preferred over all the standard CDM scenarios. As shown in Fig.~\ref{fig:density_profiles}, the inferred SIDM halo consists of a kpc-scale collisional core surrounded by a very low-density NFW-like outer halo. This corresponds to a halo mass of $M_{200}\simeq2.4 \times 10^{9}\,M_\odot$ -- roughly one order of magnitude lower than that inferred within the standard CDM framework -- though the dynamical mass remains compatible within $1\sigma$ with the value reported by \cite{Buttitta_2025}. As illustrated in Fig.~\ref{fig:velocity_field_SIDM}, the lower halo mass causes the LSB-6 velocity field to be systematically underestimated. Statistically, this yields a better overall fit, since the lower predicted velocities partially accommodate the residual outliers at the cost of larger residuals where the ordered rotational motion is followed. This behavior suggests that the statistical preference for the SIDM model should be interpreted with caution, as it may be driven by the presence of a small number of residual outliers rather than by a genuinely improved description of the kinematics.

Using Eqs.~\ref{eq:SIDM_velocity} and \ref{eq:SIDM_cross_section}, we derive the characteristic relative velocity of the SIDM particles and the corresponding velocity-weighted self-interaction cross section 
\begin{equation}
    \begin{aligned}
        \log_{10}v &= 1.30^{+0.03}_{-0.04} \,\,\,\, \mathrm{km} \, \mathrm{s}^{-1} \, , \\
        \frac{\langle \sigma v\rangle}{m} &= 14.31^{+2.01}_{-1.94} \,\,\,\, \mathrm{cm}^2 \, \mathrm{km} \, \mathrm{g}^{-1} \, \mathrm{s}^{-1} \, .
    \end{aligned}
\end{equation}
The inferred velocity-weighted cross section is consistent at the $1\sigma$ level with the empirical velocity-dependent relation of \cite{Kaplinghat:2015aga}, placing LSB-6 within the region of parameter space populated by dwarf galaxies. Rescaling this quantity yields an approximate self-interaction cross section of $\sigma/m \approx 0.7\,\mathrm{cm}^2\,\mathrm{g}^{-1}$, in excellent agreement with the values reported by \cite{Almeida:2025cee} for ultra-faint dwarf galaxies in the core-formation phase -- when only the central region of the halo has thermalized through DM self-interactions, while the outer halo still retains an NFW-like structure. On the other hand, our estimate is nearly two orders of magnitude smaller than the values reported for the gas-rich UDG AGC~114905 \citep{Nadler:2023nrd, Mancera-Pina2024}. However, it should be noted that these studies adopt a different SIDM parameterization. Our model provides an analytic description of the halo density profile and circular velocity, but does not account for heat transport or post-core-formation evolution. While this simplification may become important for systems in the core-collapse phase, it should have limited impact on our results, as the inferred properties of LSB-6 indicate that it is still in the core-formation phase.

Our LSB-6 results can also be compared with independent SIDM cross section constraints from galaxy clusters. The least stringent bound, $\sigma / m < 1.0 \,\mathrm{cm}^2 \, \mathrm{g}^{-1}$ at the $2\sigma$ level, inferred from weak-lensing measurements \citep{Adhikari:2024aff}, is fully consistent with our estimate. A more stringent $2\sigma$ upper limit of $\sigma / m < 0.613\,\mathrm{cm}^2\,\mathrm{g}^{-1}$, obtained by \cite{ODonnell:2025pkw} from a combined analysis of strong lensing and stellar kinematics data from the galaxy cluster MACS~J0138-2155, shows mild tension with our LSB-6 value. An even tighter constraint of $\sigma/m < 0.13\,\mathrm{cm}^2\,\mathrm{g}^{-1}$ at $2\sigma$, derived from strong-lensing analyses \citep{Andrade:2020lqq}, yields larger disagreement. Overall, our findings further support the hypothesis of a velocity-dependent SIDM cross section, which could reconcile the constraints inferred across different astrophysical scales.

\subsection{Non-minimally coupled dark matter}\label{susec:NMC_DM}

We also explored the possibility that the DM fluid, regardless of its microscopic nature, exhibits a NMC with gravity. To remain agnostic about the underlying DM particle, we test the NMC framework using the three standard CDM profiles considered in Sec.~\ref{sec:standardCDM_results}: a cuspy NFW profile (Eq.~\ref{NFW}), a cored Burkert profile (Eq.~\ref{Burkert}), and a soft-cored Einasto profile (Eq.~\ref{Einasto}). To ensure a comprehensive analysis, we allow the coupling sign to vary ($\epsilon=\pm1$) and perform the MCMC sampling on the parameter $L$, defined as $L^2=2\ell^2$ in the conformal NMC case and $L^2=\ell^2$ in the disformal one.

For both the NFW and Burkert profiles, the NMC framework yields halo parameters fully consistent with the standard CDM case, indicating that NMC-induced deviations are negligible over the radial range probed by our kinematic data, in agreement with previous studies \citep{Zamani:2026cxk}. This conclusion is unaffected by the sign of the coupling and is further supported by the statistical diagnostics, including both the $\chi^2$ values and the Bayes factors. For LSB-6, this equivalence also holds when the NMC framework is combined with an Einasto profile. For UDG-1, by contrast, the NMC model applied to the Einasto halo yields a substantial statistical improvement over the standard fit. However, these NMC solutions can be rejected on physical grounds.

For a positive coupling ($\epsilon=+1$), the inferred Einasto halo of UDG-1 develops an extremely extended, low-density core that extends well beyond the radial range probed by the kinematic data, making the inference strongly dependent on extrapolation. The resulting halo mass reaches $M_{200}\simeq10^{15}M_\odot$, exceeding any physically plausible value for a galaxy, while the NMC interaction length, $L\sim10^{26-38}$ kpc, is approximately $20-30$ orders of magnitude larger than the Hubble length, clearly indicating an unphysical solution.

\begin{figure}
    \centering
    \includegraphics[scale=0.3]{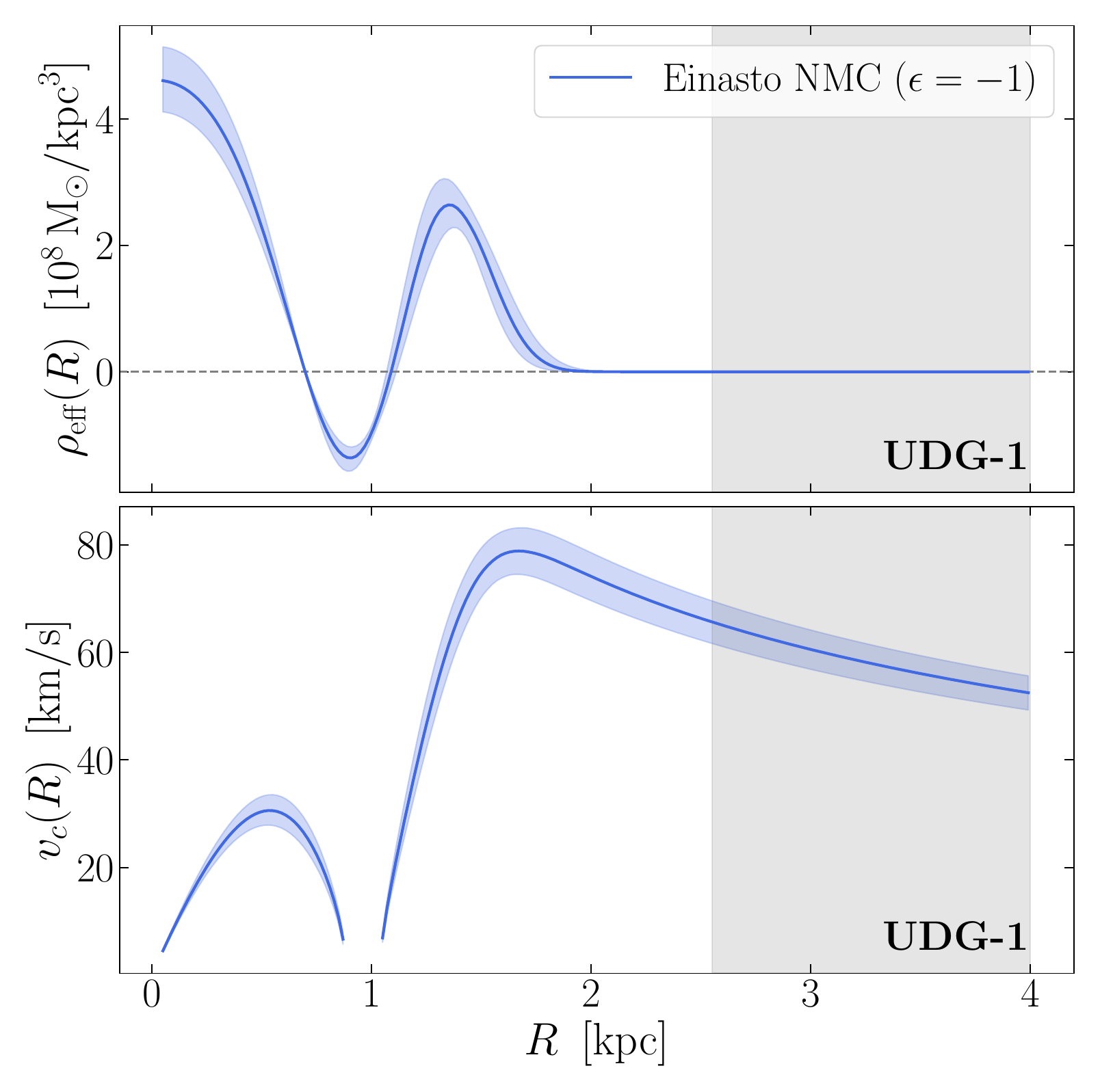}
    \caption{Effective density (\textit{top}) and velocity curve (\textit{bottom}) for the NMC model $(\epsilon = -1)$ applied to the Einasto profile of UDG-1. Solid lines show median profiles for each model, with shaded regions indicating $1\sigma$ uncertainties. The gray shaded areas mark radial ranges not covered by the observational data.} \label{fig:Einasto_NMC}
\end{figure}

For a negative coupling ($\epsilon=-1$), the inferred NMC length is $L=0.62^{+0.33}_{-0.22}$ kpc, suggesting that the NMC could contribute to the formation of a kiloparsec-scale core in the UDG-1 halo. In this case, the halo mass decreases to $M_{200}\simeq2.6\times10^{9}M_\odot$ -- nearly two orders of magnitude smaller than the standard Einasto value -- and is entirely enclosed within the effective radius of the galaxy, as indicated by the equality between $M_{200}$ and the dynamical mass. This behavior originates from the modified Poisson equation (Eq.~\ref{eq:poissonNMC}): as shown in Fig.~\ref{fig:Einasto_NMC}, the NMC contribution dominates the source term between $\sim0.5$ and $\sim1.5$ kpc, counteracting gravity and driving the source term to negative values. As a consequence, the predicted circular velocity becomes imaginary at radii around $\sim1$ kpc, making this solution physically unacceptable.

These unphysical likelihood minima arise only when the NMC framework is combined with the Einasto profile for UDG-1. This is likely a consequence of the additional flexibility provided by the Einasto parameterization, which has three free parameters and allows a lucky combination of halo and NMC parameters to substantially reduce the $\chi^2$. The imaginary circular velocities do not affect the likelihood, as the sparse spatial sampling of the UDG-1 data confines the nonphysical region of the velocity curve to a narrow radial range that is not directly probed by the observations.

Once the unphysical solutions are discarded, we derive upper limits on the NMC characteristic scale, $L$. For both LEWIS galaxies, the allowed interaction length is systematically larger for a positive coupling than for a negative one, and is generally larger for cuspless halos than for the cuspy NFW profile. For UDG-1, we obtain upper limits ranging from $\sim2\times10^{-2}$ kpc to $\sim30$ kpc, while for LSB-6 the corresponding range is slightly narrower, $\sim(4\times10^{-2},\,\,7)$ kpc. It is worth noting, however, that the inferred bound on $L$ varies significantly across cases, indicating limited sensitivity of the likelihood to this parameter. This is a consequence of the fact that, within the NMC framework, deviations from the standard Poisson equation depend not only on the amplitude of the DM density but also on its radial gradient. For halos with density profiles that are smooth over the radial range probed by the data, the Laplacian contribution is naturally suppressed, and the predicted gravitational potential approaches the $\Lambda$CDM limit.

\begin{table*}[t]
\centering
\footnotesize
\setlength{\tabcolsep}{2.5mm}
\renewcommand{\arraystretch}{1.7}

\caption{Results for UDG-1.}
\label{tab:UDG1}

\resizebox{0.9\textwidth}{!}{
\begin{tabular}{c|ccccccc|ccc}
\hline
\hline
\multicolumn{11}{c}{Stars only}\\
\hline
profile & $-$ & $-$ & $-$ & $-$ & $-$ & $\Upsilon_*$ & $\log_{10} M_{\mathrm{dyn}}$ & $\chi^2_L$ & $\chi^2_P$ & $\ln\mathcal{B}$ \\
&  & & & & & $(M_\odot / L_\odot)$ & $(M_{\odot})$ & & & \\
\hline
Maclaurin's spheroid & $-$ & $-$ & $-$ & $-$ & $-$ & $535.50^{+91.15}_{-80.37}$ & $9.20^{+0.07}_{-0.07}$ & $670.8$ & $-$ & $0.69^{+0.02}_{-0.02}$ \\
\hline
\hline
\multicolumn{11}{c}{Standard cold dark matter}\\
\hline
profile & $c_{200}$ & $\log_{10} M_{200}$ & $\gamma$ & $-$ & $-$ & $\Upsilon_*$ & $\log_{10} M_{\mathrm{dyn}}$ & $\chi^2_L$ & $\chi^2_P$ & $\ln\mathcal{B}$ \\
&  & $(M_\odot)$ & & & & $(M_\odot / L_\odot)$ & $(M_{\odot})$ & & & \\
\hline
NFW & $4.53^{+1.43}_{-1.13}$ & $11.01^{+0.47}_{-0.44}$ & $-$ & $-$ & $-$ & $1.68^{+1.13}_{-0.96}$ & $8.69^{+0.08}_{-0.09}$ & $733.3$ & $738.8$ & $-33.98^{+0.03}_{-0.02}$ \\
Burkert & $21.44^{+1.43}_{-1.47}$ & $13.43^{+0.71}_{-0.59}$ & $-$ & $-$ & $-$ & $1.65^{+1.12}_{-0.97}$ & $8.92^{+0.07}_{-0.08}$ & $675.4$ & $685.7$ & $-7.41^{+0.03}_{-0.02}$ \\ 
Einasto & $11.08^{+0.60}_{-0.65}$ & $11.69^{+1.47}_{-1.16}$ & $>1.92$ & $-$ & $-$  & $1.68^{+1.17}_{-1.00}$ & $8.98^{+0.07}_{-0.07}$ & $670.9$ & $671.1$ & $0$ \\
\hline
\hline
\multicolumn{11}{c}{Fuzzy dark matter}\\
\hline
profile & $c_{200}$ & $\log_{10} M_{200}$ & $m_\alpha$ & $-$ & $r_t$ & $\Upsilon_*$ & $\log_{10} M_{\mathrm{dyn}}$ & $\chi^2_L$ & $\chi^2_P$ & $\ln\mathcal{B}$ \\
&  & $(M_\odot)$ & $(10^{-22} \mathrm{eV})$ & & $(\mathrm{kpc})$ & $(M_\odot / L_\odot)$ & $(M_{\odot})$ & & & \\
\hline
soliton + NFW & $4.59^{+1.19}_{-0.98}$ & $10.82^{+0.41}_{-0.35}$ & $4.74^{+1.83}_{-1.14}$ & $-$ & $0.27^{+0.10}_{-0.13}$ & $1.77^{+1.12}_{-1.01}$ & $8.64^{+0.09}_{-0.10}$ & $733.2$ & $733.2$ & $-32.66^{+0.03}_{-0.02}$ \\
\hline
\hline
\multicolumn{11}{c}{Self-interacting dark matter}\\
\hline
profile & $c_{200}$ & $\log_{10} M_{200}$ & $c^{\mathrm{iso}}_{200}$ & $\log_{10} M^{\mathrm{iso}}_{200}$ & $r_t$ & $\Upsilon_*$ & $\log_{10} M_{\mathrm{dyn}}$ & $\chi^2_L$ & $\chi^2_P$ & $\ln\mathcal{B}$ \\
&  & $(M_\odot)$ & & $(M_\odot)$ & $(\mathrm{kpc})$ & $(M_\odot / L_\odot)$ & $(M_{\odot})$ & & & \\
\hline
pseudo-isoth. + NFW & $6.33^{+3.48}_{-2.62}$ & $13.22^{+1.08}_{-1.08}$ & $51.92^{+4.26}_{-4.18}$ & $21.63^{+5.73}_{-5.51}$ & $>2.37$ & $1.66^{+1.16}_{-0.98}$ & $8.98^{+0.07}_{-0.07}$ & $670.8$ & $671.0$ & $-0.07^{+0.03}_{-0.02}$ \\
\hline
\hline
\multicolumn{11}{c}{Non-minimally coupled dark matter}\\
\hline
profile & $c_{200}$ & $\log_{10} M_{200}$ & $\gamma$ & $\log_{10}L$ & $-$ & $\Upsilon_*$ & $\log_{10} M_{\mathrm{dyn}}$ & $\chi^2_L$ & $\chi^2_P$ & $\ln\mathcal{B}$ \\
&  & $(M_\odot)$ & $(\mathrm{kpc})$ & & & $(M_\odot / L_\odot)$ & $(M_{\odot})$ & & & \\
\hline
NFW $(\epsilon=+1)$ & $4.50^{+1.33}_{-1.12}$ & $11.02^{+0.47}_{-0.42}$ & $-$ & $<0.71$ & $-$ & $1.78^{+1.17}_{-1.02}$ & $8.69^{+0.08}_{-0.09}$ & $733.2$ & $738.8$ & $-34.00^{+0.02}_{-0.03}$ \\
NFW $(\epsilon=-1)$ & $4.54^{+1.31}_{-1.12}$ & $11.03^{+0.45}_{-0.43}$ & $-$ & $<-1.69$ & $-$ & $1.76^{+1.12}_{-1.03}$ & $8.70^{+0.08}_{-0.09}$ & $733.3$ & $738.8$ & $-33.98^{+0.02}_{-0.02}$ \\
Burkert $(\epsilon=+1)$ & $21.41^{+1.47}_{-1.46}$ & $13.42^{+0.70}_{-0.60}$ & $-$ & $<1.48$ & $-$ & $1.71^{+1.12}_{-0.99}$ & $8.92^{+0.08}_{-0.08}$ & $675.4$ & $685.7$ & $-7.42^{+0.02}_{-0.03}$ \\ 
Burkert $(\epsilon=-1)$ & $21.43^{+1.43}_{-1.44}$ & $13.43^{+0.73}_{-0.62}$ & $-$ & $<0.43$ & $-$ & $1.66^{+1.10}_{-0.95}$ & $8.92^{+0.07}_{-0.08}$ & $675.4$ & $685.7$ & $-7.39^{+0.02}_{-0.02}$ \\ 
Einasto $(\epsilon=+1)$ & $4.56^{+0.99}_{-1.01}$ & $15.04^{+1.43}_{-1.48}$ & $31.52^{+5.62}_{-4.56}$ & $36.76^{+11.32}_{-10.60}$ & $-$ & $1.68^{+1.15}_{-1.01}$ & $7.93^{+0.24}_{-0.29}$ & $560.4$ & $560.5$ & $73.60^{+0.03}_{-0.03}$ \\ 
Einasto $(\epsilon=-1)$ & $29.08^{+1.08}_{-1.08}$ & $9.41^{+0.05}_{-0.05}$ & $4.38^{+0.22}_{-0.19}$ & $-0.21^{+0.19}_{-0.19}$ & $-$ & $1.70^{+1.16}_{-1.02}$ & $9.41^{+0.05}_{-0.05}$ & $546.9$ & $551.0$ & $76.43^{+0.05}_{-0.04}$ \\ 
\hline
\hline
\end{tabular}}
\tablefoot{The quantity $\chi^2_L$ refers to the likelihood-only contribution, excluding any priors, whereas $\chi^2_P$ includes the effect of the priors. The Bayes factor is denoted by $\ln \mathcal{B}$ . For the fuzzy and self-interacting dark matter models, the parameter $\log_{10} M_{200}$ is not directly sampled and is instead reported as a derived quantity representing the halo mass. The dynamical mass within the effective radius, $\log_{10} M_{\mathrm{dyn}}$, is also reported as a derived quantity.}
\end{table*}

\begin{table*}[t]
\centering
\footnotesize
\setlength{\tabcolsep}{2.5mm}
\renewcommand{\arraystretch}{1.7}

\caption{Results for LSB-6.}
\label{tab:LSB6}

\resizebox{0.9\textwidth}{!}{
\begin{tabular}{c|ccccccc|ccc}
\hline
\hline
\multicolumn{11}{c}{Stars only}\\
\hline
profile & $-$ & $-$ & $-$ & $-$ & $-$ & $\Upsilon_*$ & $\log_{10} M_{\mathrm{dyn}}$ & $\chi^2_L$ & $\chi^2_P$ & $\ln\mathcal{B}$ \\
&  & & & & & $(M_\odot / L_\odot)$ & $(M_{\odot})$ & & & \\
\hline
Maclaurin's spheroid & $-$ & $-$ & $-$ & $-$ & $-$ & $321.61^{+17.78}_{-17.54}$ & $10.02^{+0.02}_{-0.02}$ & $1954.8$ & $-$ & $-37.96^{+0.02}_{-0.03}$ \\
\hline
\hline
\multicolumn{11}{c}{Standard cold dark matter}\\
\hline
profile & $c_{200}$ & $\log_{10} M_{200}$ & $\gamma$ & $-$ & $-$ & $\Upsilon_*$ & $\log_{10} M_{\mathrm{dyn}}$ & $\chi^2_L$ & $\chi^2_P$ & $\ln\mathcal{B}$ \\
&  & $(M_\odot)$ & & & & $(M_\odot / L_\odot)$ & $(M_{\odot})$ & & & \\
\hline
NFW & $3.72^{+0.96}_{-0.84}$ & $11.72^{+0.37}_{-0.34}$ & $-$ & $-$ & $-$ & $3.15^{+0.60}_{-0.60}$ & $9.39^{+0.03}_{-0.03}$ & $1922.9$ & $1929.3$ & $-25.97^{+0.02}_{-0.03}$ \\
Burkert & $31.41^{+1.40}_{-1.29}$ & $10.98^{+0.12}_{-0.11}$ & $-$ & $-$ & $-$ & $3.12^{+0.59}_{-0.61}$ & $9.43^{+0.05}_{-0.05}$ & $1876.8$ & $1885.2$ & $-3.90^{+0.02}_{-0.03}$ \\
Einasto & $12.20^{+0.98}_{-1.71}$ & $9.81^{+0.54}_{-0.30}$ & $0.88^{+0.36}_{-0.29}$ & $-$ & $-$ & $3.08^{+0.61}_{-0.56}$ & $9.45^{+0.07}_{-0.07}$ & $1876.6$ & $1876.6$ & $0$ \\
\hline
\hline
\multicolumn{11}{c}{Fuzzy dark matter}\\
\hline
profile & $c_{200}$ & $\log_{10} M_{200}$ & $m_\alpha$ & $-$ & $r_t$ & $\Upsilon_*$ & $\log_{10} M_{\mathrm{dyn}}$ & $\chi^2_L$ & $\chi^2_P$ & $\ln\mathcal{B}$ \\
&  & $(M_\odot)$ & $(10^{-22} \mathrm{eV})$ & & $(\mathrm{kpc})$ & $(M_\odot / L_\odot)$ & $(M_{\odot})$ & & & \\
\hline
soliton + NFW & $3.90^{+0.97}_{-0.89}$ & $11.64^{+0.38}_{-0.33}$ & $6.26^{+0.41}_{-0.38}$ & $-$ & $0.18^{+0.02}_{-0.03}$ & $3.14^{+0.62}_{-0.60}$ & $9.39^{+0.03}_{-0.03}$ & $1913.5$ & $1913.6$ & $-21.09^{+0.02}_{-0.02}$ \\
\hline
\hline
\multicolumn{11}{c}{Self-interacting dark matter}\\
\hline
profile & $c_{200}$ & $\log_{10} M_{200}$ & $c^{\mathrm{iso}}_{200}$ & $\log_{10} M^{\mathrm{iso}}_{200}$ & $r_t$ & $\Upsilon_*$ & $\log_{10} M_{\mathrm{dyn}}$ & $\chi^2_L$ & $\chi^2_P$ & $\ln\mathcal{B}$ \\
&  & $(M_\odot)$ & & $(M_\odot)$ & $(\mathrm{kpc})$ & $(M_\odot / L_\odot)$ & $(M_{\odot})$ & & & \\
\hline
pseudo-isoth. + NFW & $26.82^{+6.20}_{-5.70}$ & $9.38^{+0.12}_{-0.11}$ & $61.01^{+8.01}_{-3.66}$ & $>10.08$ & $0.75^{+0.07}_{-0.08}$ & $2.98^{+0.61}_{-0.61}$ & $8.91^{+0.06}_{-0.07}$ & $1673.7$ & $1680.3$ & $98.20^{+0.03}_{-0.03}$ \\
\hline
\hline
\multicolumn{11}{c}{Non-minimally coupled dark matter}\\
\hline
profile & $c_{200}$ & $\log_{10} M_{200}$ & $\gamma$ & $\log_{10}L$ & $-$ & $\Upsilon_*$ & $\log_{10} M_{\mathrm{dyn}}$ & $\chi^2_L$ & $\chi^2_P$ & $\ln\mathcal{B}$ \\
&  & $(M_\odot)$ & $(\mathrm{kpc})$ & & & $(M_\odot / L_\odot)$ & $(M_{\odot})$ & & & \\
\hline
NFW $(\epsilon=+1)$ & $3.77^{+0.97}_{-0.87}$ & $11.70^{+0.37}_{-0.34}$ & $-$ & $<0.43$ & $-$ & $3.16^{+0.62}_{-0.61}$ & $9.39^{+0.03}_{-0.03}$ & $1923.0$ & $1929.3$ & $-25.96^{+0.03}_{-0.02}$ \\
NFW $(\epsilon=-1)$ & $3.78^{+0.95}_{-0.88}$ & $11.70^{+0.38}_{-0.33}$ & $-$ & $<-1.44$ & $-$ & $3.15^{+0.60}_{-0.61}$ & $9.39^{+0.03}_{-0.03}$ & $1923.0$ & $1929.3$ & $-25.96^{+0.04}_{-0.03}$ \\
Burkert $(\epsilon=+1)$ & $31.42^{+1.40}_{-1.32}$ & $10.98^{+0.12}_{-0.11}$ & $-$ & $<0.82$ & $-$ & $3.08^{+0.62}_{-0.60}$ & $9.43^{+0.05}_{-0.05}$ & $1876.8$ & $1885.2$ & $-3.92^{+0.02}_{-0.03}$ \\ 
Burkert $(\epsilon=-1)$ & $31.42^{+1.43}_{-1.37}$ & $10.98^{+0.12}_{-0.12}$ & $-$ & $<0.62$ & $-$ & $3.10^{+0.61}_{-0.59}$ & $9.43^{+0.05}_{-0.05}$ & $1876.8$ & $1885.2$ & $-3.92^{+0.02}_{-0.03}$ \\ 
Einasto $(\epsilon=+1)$ & $12.00^{+1.11}_{-2.25}$ & $9.74^{+0.55}_{-0.25}$ & $0.93^{+0.57}_{-0.32}$ & $<0.82$ & $-$ & $3.00^{+0.64}_{-0.65}$ & $9.44^{+0.07}_{-0.09}$ & $1878.6$ & $1878.7$ & $-0.15^{+0.03}_{-0.03}$ \\
Einasto $(\epsilon=-1)$ & $12.22^{+0.98}_{-1.73}$ & $9.80^{+0.54}_{-0.30}$ & $0.88^{+0.36}_{-0.29}$ & $<-0.09$ & $-$ & $3.15^{+0.58}_{-0.59}$ & $9.45^{+0.07}_{-0.07}$ & $1876.6$ & $1876.6$ & $0.00^{+0.03}_{-0.03}$ \\ 
\hline
\hline
\end{tabular}}
\tablefoot{Same as Table~\ref{tab:UDG1} but for LSB-6.}
\end{table*}

\section{Dark matter constraints on formation scenario of UDGs and LSBs}
\label{sec:formation_UDGs}

The DM content is a key discriminant among the various formation channels proposed for UDGs. \citealt{vanDokkum2015} first proposed that UDGs might be failed $L^\ast$ galaxies with Milky Way-like halo masses ($M_{h}\sim10^{12}\,M_\odot$) that lost their gas supply at early epochs and subsequently evolved passively into UDGs after star formation ceased. Alternatively, UDGs may originate from dwarf galaxies with typical halo masses of $M_{h}\sim10^{10-11}\,M_\odot$, whose stellar distribution has been puffed up by internal or environmental processes \citep{Amorisco2016, Rong2017, Cintio2017, Tremmel2020}. A third possibility involves tidal UDGs, formed either from collisional debris during galaxy mergers \citep{Lelli2015, Duc2015} or from gas clumps in ram-pressure tails of galaxies falling into clusters \citep{Poggianti2019}. Such systems were born from the gravitational collapse of stripped material from parent galaxies and are thus expected to be strongly DM-deficient or nearly DM-free. The globular cluster (GC) counts provide additional constraints on the formation channel. Ultra-diffuse galaxies, in fact, display diverse GC populations: some are GC-rich beyond expectations for their stellar mass, suggesting overmassive DM halos, while others have GC systems indistinguishable from those of typical dwarf galaxies \citep{BeasleyTrujillo2016, Amorisco2018, Lim2020, Saifollahi2021}. 

The results we obtained from the different tested models, regardless of the assumed microscopic nature of the DM, suggest that both UDG-1 and LSB-6 host a high DM amount ($\sim10^{9-11}\,M_\odot$), comparable with halos of typical dwarf galaxies with similar stellar masses \citep{Gannon2026}. These findings rule out the failed $L^\ast$ galaxy progenitor scenario, which would require a significantly more massive DM halo, and disfavor a tidal, DM-free origin. The spectroscopic analysis of all compact sources in the field of view of UDG-1 and LSB-6 revealed one bound GC for UDG-1 and none for LSB-6 \citep{Mirabile2026b_submitted}. Fainter GC candidates, with S/N too low ($\lesssim2.5 $\AA$^{-1}$) for spectral analysis, were studied using multiband imaging and statistical background decontamination \citep{DAbrusco2016, Cantiello2020, Mirabile2024}, showing that neither UDG-1 nor LSB-6 exhibits statistically significant GC overdensity. The low number of spectroscopically confirmed GCs in LSB-6 and the absence of significant GC overdensity in UDG-1 prevented us from obtaining an independent dynamical mass estimate \citep{Burkert2020} but favor a puffed-up dwarf progenitor scenario over that of a typical GC-rich system from a luminous progenitor.

Considering additional structural properties, UDG-1 and LSB-6 show striking differences. UDG-1 is located in the innermost region of the Hydra I cluster -- classified as an early infaller -- where strong tidal forces might have played a role in shaping the properties and the evolution of this galaxy \citep{Buttitta_2025}. However, its morphology appears undisturbed and elongated, without a clear central light concentration, as expected in diffuse systems such as UDGs. The analysis of the stellar population showed that UDG-1 is characterized by a peculiar star-formation history which consists of an initial efficient star-formation activity, building up 50\% of its stellar mass at $t_{50}\sim13.6$ Gyr, with a more recent secondary star-formation episode \citep{Doll2026}. Among the analyzed UDGs in the LEWIS sample, the metallicity of the stellar population in UDG-1 ([M/H] $\sim -0.5$ dex) is higher than the value of [M/H] in other UDGs, which have metallicity consistent with the mass-metallicity relation for dwarf galaxies \citep{Kirby2013}. LSB-6, instead, is located at a higher clustercentric distance -- classified as a late infaller --  and its photometric characteristics are similar to those of LSB galaxies, i.e., an enhanced brighter central region embedded in a more extended and diffuse light component. Spectroscopic analysis revealed dwarf-like metallicity ([M/H]$\sim-1$ dex) and a dwarf-like star-formation history, gradually building 50\% of its stellar mass at $t_{50}\sim9$ Gyr \citep{Doll2026}.

Despite these differences, UDG-1 and LSB-6 have remarkable similarities in terms of internal dynamics, and their DM content is consistent with dwarf-like halos. We argue that both UDG-1 and LSB-6 are consistent with a puffed-up dwarf formation scenario, although the involved mechanisms could be different. Taking all the properties together, the morphology and peculiar star formation history of UDG-1 suggest that environmental forces have not been efficiently quenching the evolution of the progenitor. Internal mechanisms such as a high-spin DM halo or strong star-formation feedback might instead be responsible for puffing up the stellar distribution of the progenitor, turning a dwarf galaxy into a UDG. On the other hand, structural properties of LSB-6 are consistent with typical properties of dwarf galaxies. The progenitor of LSB-6 might be a dwarf galaxy and an external perturbation from a nearby source might be responsible for inflating its structure and inducing a twist in the most external isophotes in LSB-6, without appreciably altering its internal stellar kinematics. To conclude, UDG-1 and LSB-6 likely share a similar formation pathway despite their different present-day structural properties, and demonstrate that DM content alone cannot fully discriminate between the specific mechanisms driving the formation of individual faint and diffuse systems.

\section{Conclusions}\label{sec:conclusions}

We investigated the dynamics of two extreme LSB galaxies in the Hydra I cluster, UDG-1 and LSB-6, by modeling the stellar velocity fields derived from IF MUSE spectroscopy within the LEWIS project. Both galaxies were modeled as rotating spheroidal systems with typical dwarf-like thickness, embedded in spherical DM halos. We explored the nature of the DM particle within four theoretical frameworks: standard CDM, FDM, SIDM, and NMC-DM. Our main findings can be summarized as follows:
\begin{itemize}
    \item the DM-free scenario can be robustly excluded for both galaxies;
    \item within the standard CDM framework, our analysis, together with the comparison to theoretical relations and simulated halo populations, indicates that a cuspy halo provides the most physically plausible description of UDG-1, whereas a cuspless halo is favored for LSB-6;
    \item FDM emerges as a viable alternative, yielding mutually consistent constraints on the ALP mass for UDG-1 and LSB-6;
    \item the SIDM results are strongly affected by residual outliers in the kinematic data. The LSB-6 results nonetheless appear robust, yielding constraints on the self-interaction cross section consistent with the literature;
    \item the NMC-DM framework remains viable provided that deviations from GR are minimal.
\end{itemize}
For both galaxies, regardless of the assumed DM model, the dynamical masses are in excellent agreement with the estimates of \cite{Buttitta_2025}, obtained with an independent method.

This study represents the first in a series aimed at exploiting the kinematics of extreme LSB galaxies as benchmarks to probe galaxy dynamics, DM physics, and the nature of gravity. As a pilot project, we adopted a realistic yet simplified modeling of the galaxies, particularly their stellar component, and found that the results are robust to these assumptions at the current data quality, with different stellar models affecting the inferred dynamics by $\lesssim 10\%$. With more extended and higher-quality datasets, however, more accurate stellar modeling will be required to minimize biases and properly account for the intrinsic differences between UDGs and extended dwarf galaxies.

For this analysis, UDG-1 and LSB-6 were selected from the LEWIS sample because of their regular morphology and coherent rotation. Five additional galaxies with available stellar velocity fields, but more complex morphology and dynamics, will be examined in future work. In parallel, complementary tracers will be incorporated to provide a more complete dynamical picture. In particular, ionized-gas kinematics will be used to investigate a peculiar, star-forming, gas-rich UDG in the Hydra I cluster \citep{Rossi2026arXiv}.

From a theoretical perspective, future work will extend the present analysis to a broader class of exotic DM candidates and alternative gravity frameworks to test the robustness of our conclusions across fundamentally different paradigms. By systematically combining high-quality IF spectroscopy with detailed dynamical modeling, this series will consolidate extreme LSB galaxies as precision laboratories for fundamental physics.

\begin{acknowledgements}
We wish to thank the anonymous Referee whose comments helped us to improve the clarity of the manuscript. This work is based on observations collected at the European Southern Observatory under ESO programs 108.222P.001, 108.222P.002, 108.222P.003. F.B. and  S.C. acknowledge the support of Istituto Nazionale di Fisica Nucleare (INFN), iniziativa specifica QGSKY. The research of S.Z. and V.S. is funded by the Polish National Science Centre grant No. DEC-2021/43/O/ST9/00664. E.I. acknowledges support by the INAF GO funding grant 2022-2023. E.I. acknowledges the support by the Italian Ministry for Education University and Research (MIUR) grant PRIN 2022 2022383WFT “SUNRISE”, CUP C53D23000850006. M. Miranda acknowledges the support of INFN, iniziativa specifica MOONLIGHT-2. E.M.C. is funded by the grants DOR 2023-2026 of the University of Padova. This publication is based upon work from COST Action CA21136 -- ``Addressing observational tensions in cosmology with systematics and fundamental physics (CosmoVerse)'', supported by COST (European Cooperation in Science and Technology).
\end{acknowledgements}

\bibliographystyle{aa}
\bibliography{biblio}

\begin{appendix}

\section{Additional cold dark matter profiles}\label{Appendix_CDM}

\begin{table*}[b]
\centering
\footnotesize
\setlength{\tabcolsep}{2.5mm}
\renewcommand{\arraystretch}{1.4}

\vspace{0.3cm}
\caption{Results for UDG-1 (upper table) and LSB-6 (lower table).}\label{tab:appendixCDM}

\resizebox{0.75\textwidth}{!}{
\begin{tabular}{c|ccccc|ccc}
\hline
\hline
\multicolumn{9}{c}{Standard Cold Dark Matter}\\
\hline
profile & $c_{200}$ & $\log_{10} M_{200}$ & $\gamma$ & $\Upsilon_*$ & $\log_{10} M_{\mathrm{dyn}}$ & $\chi^2_L$ & $\chi^2_P$ & $\ln\mathcal{B}$ \\
& & $(M_\odot)$ & & $(M_\odot / L_\odot)$ & $(M_\odot)$ & & & \\
\hline
Hernquist & $4.68^{+1.38}_{-1.17}$ & $11.02^{+0.52}_{-0.49}$ & $-$ & $1.69^{+1.13}_{-0.98}$ & $8.70^{+0.08}_{-0.09}$ & $732.1$ & $737.2$ & $-33.20^{+0.02}_{-0.02}$ \\
gNFW & $7.49^{+0.65}_{-0.76}$ & $14.65^{+1.01}_{-0.87}$ & $<0.04$ & $1.65^{+1.17}_{-1.00}$ & $8.95^{+0.07}_{-0.07}$ & $674.6$ & $676.0$ & $-2.71^{+0.03}_{-0.02}$ \\
DARKexp & $8.98^{+0.64}_{-0.74}$ & $14.23^{+0.96}_{-0.86}$ & $<0.04$ & $1.66^{+1.08}_{-0.96}$ & $8.95^{+0.07}_{-0.08}$ & $674.9$ & $676.8$ & $-3.01^{+0.03}_{-0.03}$ \\
\hline
\hline
\noalign{\vskip 0.3cm}
\hline
\hline
\multicolumn{9}{c}{Standard Cold Dark Matter}\\
\hline
profile & $c_{200}$ & $\log_{10} M_{200}$ & $\gamma$ & $\Upsilon_*$ & $\log_{10} M_{\mathrm{dyn}}$ & $\chi^2_L$ & $\chi^2_P$ & $\ln\mathcal{B}$ \\
& & $(M_\odot)$ & & $(M_\odot / L_\odot)$ & $(M_\odot)$ & & & \\
\hline
Hernquist & $4.00^{+1.04}_{-0.97}$ & $11.71^{+0.44}_{-0.39}$ & $-$ & $3.16^{+0.60}_{-0.60}$ & $9.40^{+0.03}_{-0.03}$ & $1921.9$ & $1927.4$ & $-25.00^{+0.03}_{-0.03}$ \\
gNFW & $11.78^{+1.10}_{-1.53}$ & $10.79^{+0.28}_{-0.19}$ & $<0.17$ & $3.10^{+0.59}_{-0.60}$ & $9.49^{+0.05}_{-0.05}$ & $1877.4$ & $1877.4$ & $-0.08^{+0.03}_{-0.03}$ \\
DARKexp & $12.99^{+1.00}_{-1.56}$ & $10.52^{+0.33}_{-0.20}$ & $<0.18$ & $3.10^{+0.60}_{-0.61}$ & $9.49^{+0.05}_{-0.05}$ & $1877.4$ & $1877.5$ & $-0.11^{+0.03}_{-0.04}$ \\
\hline
\hline
\end{tabular}}
\tablefoot{The quantity $\chi^2_L$ refers to the likelihood-only contribution, excluding any priors, whereas $\chi^2_P$ includes the effect of the priors. $\ln \mathcal{B}$ denotes the Bayes factor. The dynamical mass within the effective radius, $\log_{10} M_{\mathrm{dyn}}$, is not directly sampled and is instead reported as a derived quantity.}
\end{table*}

Within the standard CDM framework, which we adopt as our reference, we also consider additional DM density profiles beyond the NFW (Eq.~\ref{NFW}), Burkert (Eq.~\ref{Burkert}), and Einasto (Eq.~\ref{Einasto}) forms. This allows us to verify that our analysis and model comparison are not biased by the specific choice of these commonly used parameterizations. In the following, we provide a brief overview of the considered density models and discuss the results of our dynamical analysis for UDG-1 and LSB-6, listed in Table~\ref{tab:appendixCDM}.

We first consider the Hernquist profile \citep{Hernquist_profile}, defined as
\begin{equation}\label{Hernquist}
    \rho_{\mathrm{H}}(r) = \rho_s \bigg(\frac{r}{r_s}\bigg)^{-1} \bigg( 1 + \frac{r}{r_s} \bigg)^{- 3} \, ,
\end{equation}
where
\begin{equation}
    \rho_s = \frac{200}{3} \rho_c \, c_{200} \left( 1 + \frac{c_{200}}{2} \right)^{2} \, ,
\end{equation}
and the radius at which the logarithmic slope equals $-2$ is given by $r_{-2} = r_s / 2$. This analytic profile was originally introduced to reproduce the de Vaucouleurs law for the surface brightness of elliptical galaxies \citep{1948AnAp...11..247D}, and it is frequently employed to model dSph systems. Similarly to the NFW profile, the Hernquist density distribution scales as $r^{-1}$ in the inner regions, thus exhibiting a central cusp. As a consequence, the constraints derived from the dynamical analysis of UDG-1 and LSB-6 are effectively equivalent to those obtained with the NFW profile.

We next examine a generalized NFW (gNFW) profile \citep{Jing:1999ir,Graham:2005xx}, which can be expressed as
\begin{equation}\label{gNFW}
    \rho_{\mathrm{gNFW}}(r) = \rho_s \bigg(\frac{r}{r_s}\bigg)^{-\gamma} \bigg( 1 + \frac{r}{r_s} \bigg)^{\gamma - 3} \, ,
\end{equation}
where the additional free parameter $\gamma \in [0,2]$ controls the inner slope of the density profile. In particular, $\gamma = 0$ corresponds to a cored halo, while $\gamma > 0$ yields a cuspy distribution that reduces to the standard NFW form for $\gamma = 1$. The characteristic density and scale radius are given by
\begin{equation}
    \rho_s = \frac{200}{3} \rho_c \, c_{200}^\gamma \frac{\left( 3 - \gamma \right) \left( 2 - \gamma \right)^\gamma}{{}_2F_1\left[ 3-\gamma, 3-\gamma, 4-\gamma, \left( \gamma - 2 \right) c_{200} \right]}  \, ,
\end{equation}
and $r_{-2} = (2-\gamma) r_s$. The gNFW profile is commonly adopted to introduce additional flexibility in halo modeling and to capture the diversity observed in galaxy rotation curves. In our analysis, it reproduces the kinematics of both systems with good statistical agreement. For both UDG-1 and LSB-6, the gNFW model performs comparably to the cuspless profiles -- the reference soft-cored Einasto model and the cored Burkert profile -- and better than the cuspy NFW model. However, the gNFW model does not alleviate the issue of the unrealistically large halo mass in the case of UDG-1. For LSB-6, we derive an upper limit on the inner-slope parameter, finding $\gamma < 0.17$, with the posterior distribution peaking toward $\gamma \rightarrow 0$. The dynamical analysis within the gNFW framework therefore reinforces the conclusion that the spectroscopic observations of LSB-6 favor cored DM density profiles.

We also consider an additional flexible density profile, originally introduced in \citet{Hjorth_2010} and \citet{Williams_2010}. The DARKexp profile is derived from equilibrium statistical-mechanical arguments applied to collisionless, self-gravitating systems with isotropic velocity distributions. It can be written as
\begin{equation}\label{DARKexp}
    \rho_{\mathrm{DARK}}(r) = \rho_s \bigg(\frac{r}{r_s}\bigg)^{-\gamma} \bigg( 1 + \frac{r}{r_s} \bigg)^{\gamma - 4} \, ,
\end{equation}
where 
\begin{equation}
    \rho_s = \frac{200}{3} \rho_c \, c_{200}^3 \left( 1 - \frac{\gamma}{2} \right)^3 \left( 3 - \gamma \right) \frac{1 + \left( 1 - \frac{\gamma}{2} \right) c_{200}}{\left( 1 - \frac{\gamma}{2} \right) c_{200}}  \, ,
\end{equation}
and $r_{-2} = (1-\gamma/2) r_s$. As in the gNFW case, the inner slope -- and hence the cuspiness of the halo -- is regulated by the parameter $\gamma$. However, Eq.~\ref{DARKexp} never reduces exactly to an NFW profile: it can only mimic the NFW behavior in the innermost regions, while exhibiting a smoother transition toward the outer halo. Owing to this flexibility, the DARKexp profile is well suited to capture the non-universality of halo density shapes and has been successfully applied to the modeling of dwarf galaxies \citep{Hjorth:2015bfa}. Our dynamical analysis of UDG-1 and LSB-6 yields constraints that are largely consistent with those obtained using the gNFW profile.

Our supplementary analysis confirms the results presented in the main body of the paper.

\section{Posterior distributions}\label{AppendixB:fig_posteriors}

In this appendix we present the posterior distributions obtained from our Bayesian analysis of the kinematics of UDG-1 and LSB-6. The posterior distribution for the DM-free case is not shown, as the sampled parameter space is one-dimensional ($\Upsilon_*$), making the dynamical mass fully correlated with the sole free parameter.

\begin{figure*}[b]
    \centering
    \vspace{0.3cm}
    \includegraphics[scale=0.6]{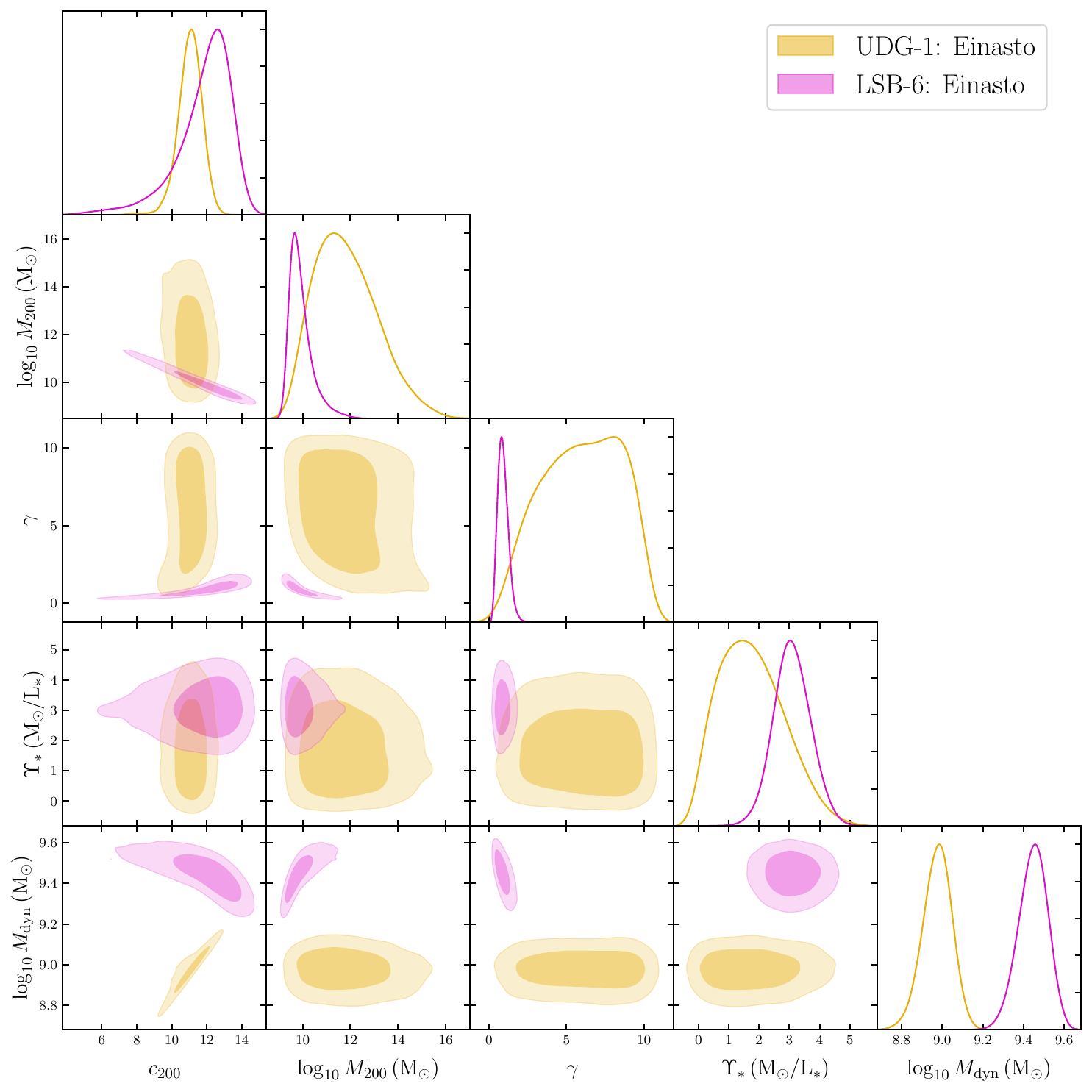}
    \caption{UDG-1 and LSB-6 posterior distributions. Standard cold dark matter model: soft-cored Einasto profile. The dynamical mass $\log_{10} M_{\mathrm{dyn}}$ is a derived parameter.}\label{fig:posteriors_Einasto}
    \vspace{4cm}
\end{figure*}

\begin{figure*}
    \centering
    \includegraphics[scale=0.57]{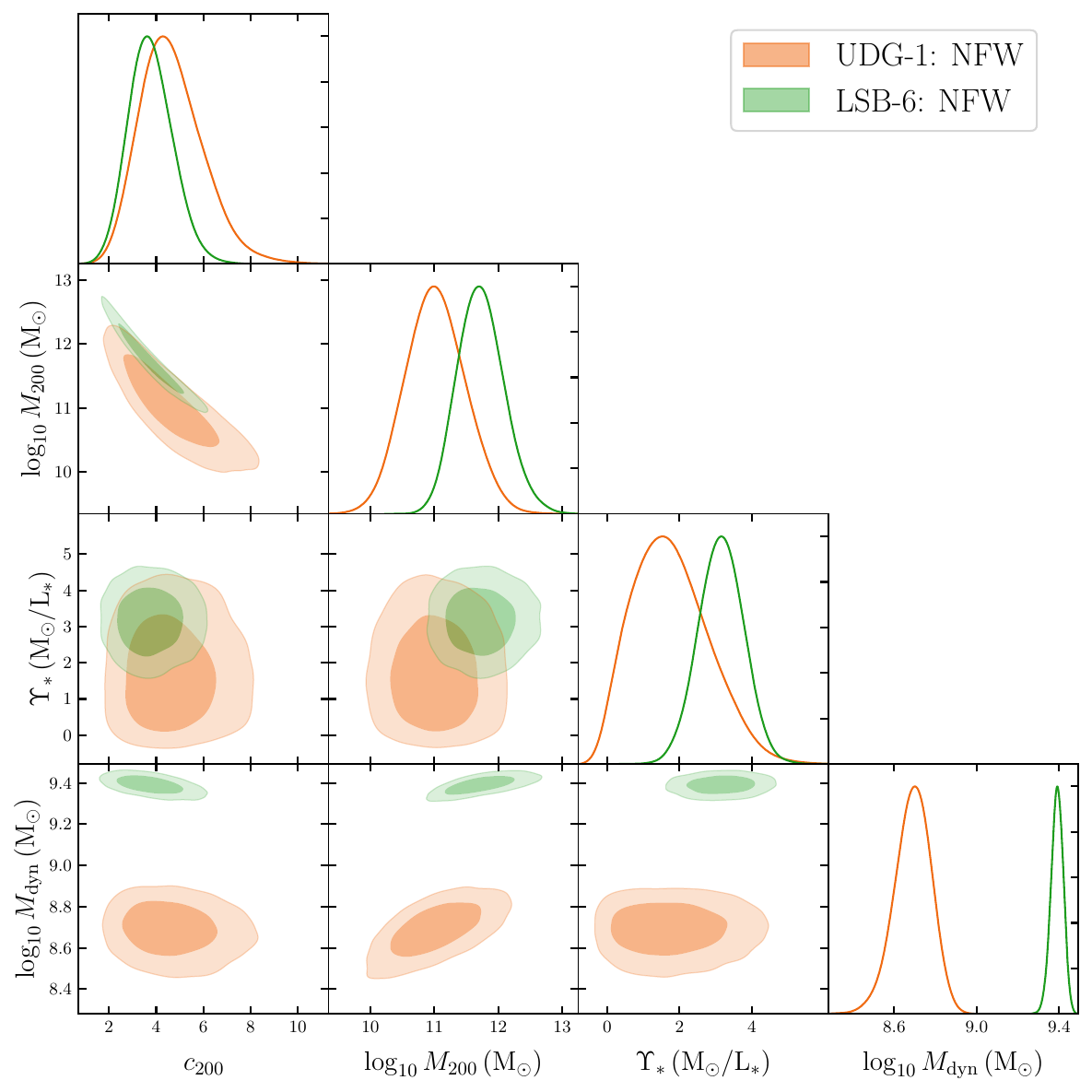} \\[0.5cm]
    \includegraphics[scale=0.57]{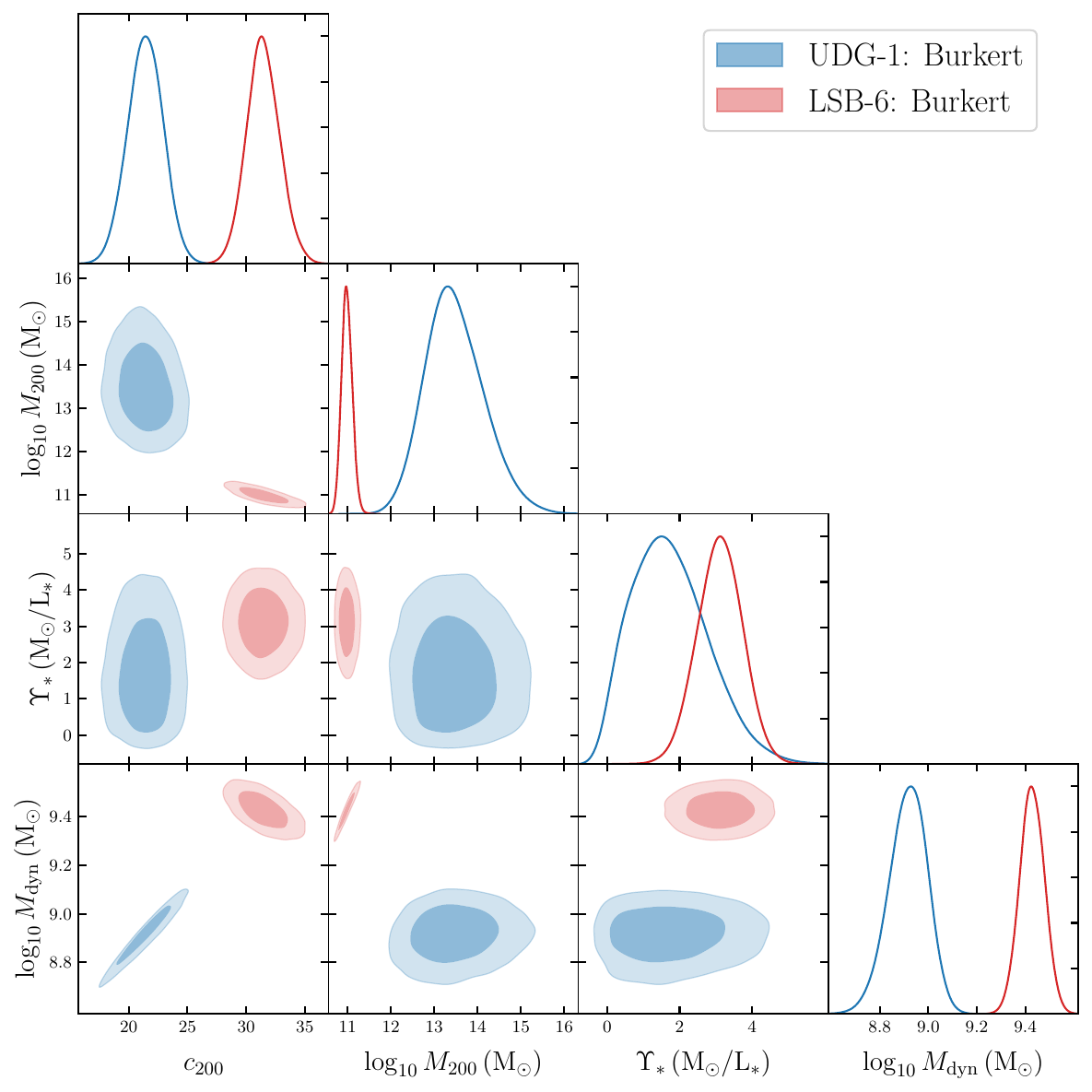} 
    \caption{UDG-1 and LSB-6 posterior distributions. Standard cold dark matter models: cuspy NFW (top) and cored Burkert (bottom) profiles. The dynamical mass $\log_{10} M_{\mathrm{dyn}}$ is a derived parameter.}\label{fig:posteriors_NFW-Burkert}
\end{figure*}

\begin{figure*}
    \centering
    \includegraphics[scale=0.57]{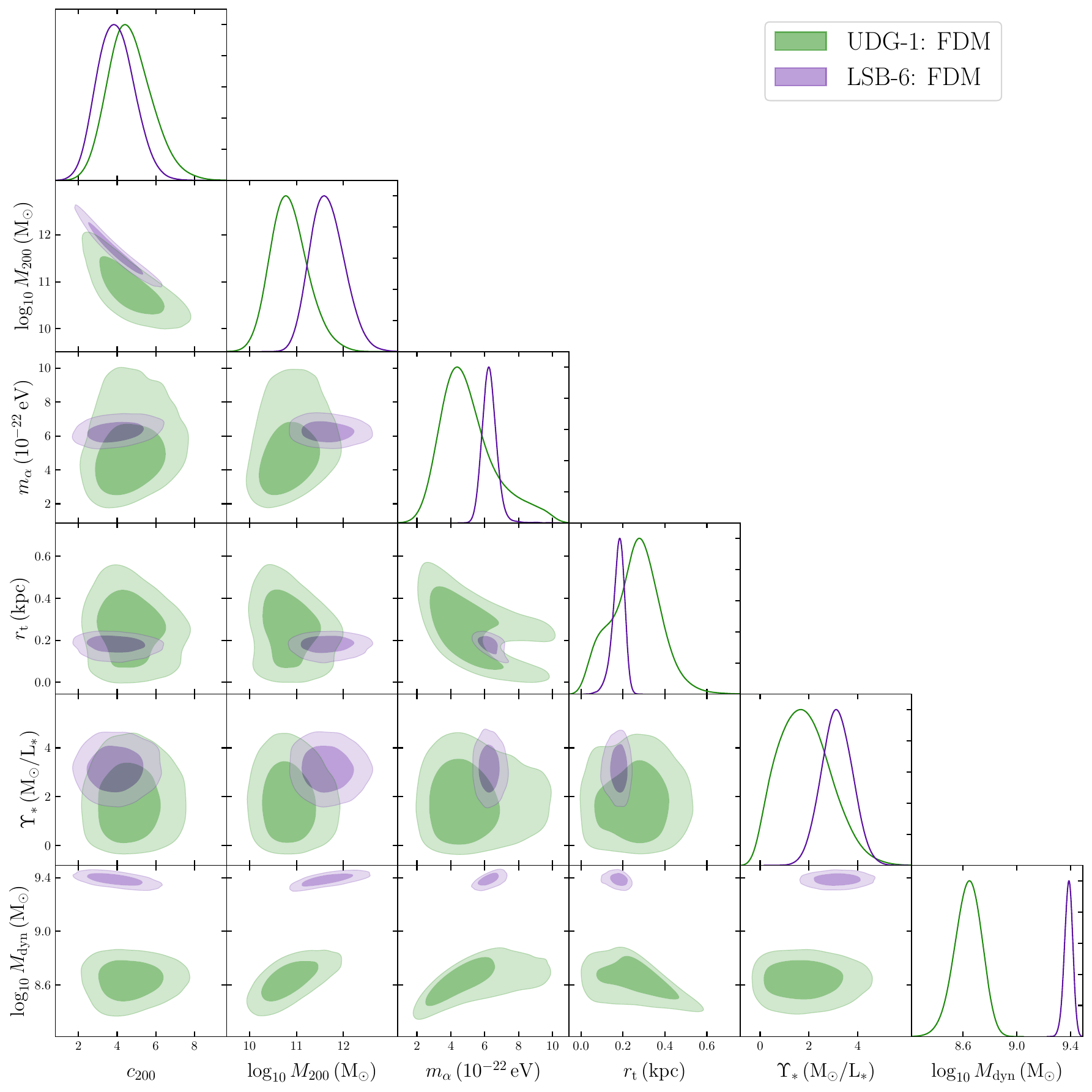}
    \caption{UDG-1 and LSB-6 posterior distributions. Fuzzy dark matter model. The halo mass $\log_{10} M_{200}$ and the dynamical mass $\log_{10} M_{\mathrm{dyn}}$ are derived parameters.}\label{fig:posteriors_FDM}
\end{figure*}

\begin{figure*}
    \centering
    \includegraphics[scale=0.45]{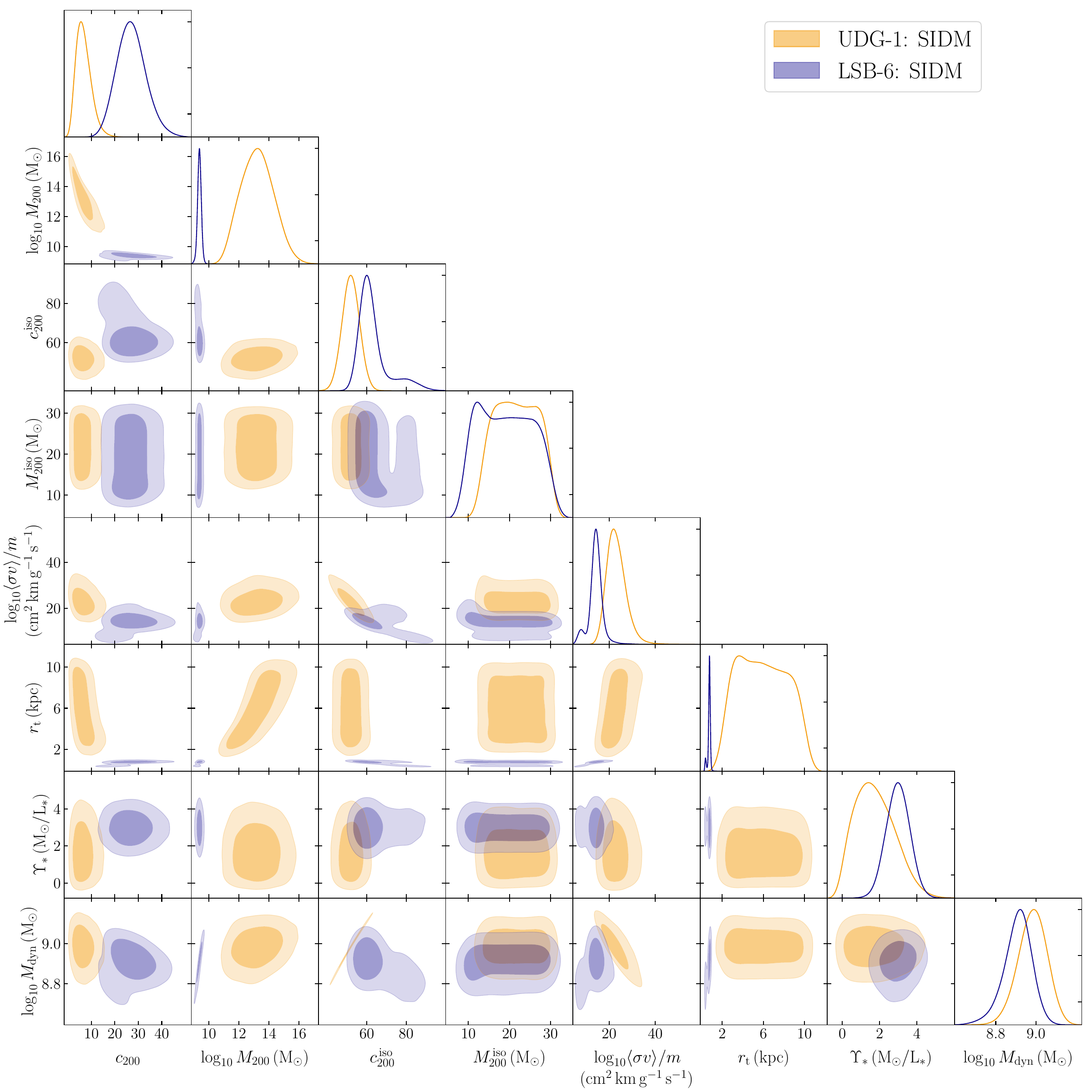}
    \caption{UDG-1 and LSB-6 posterior distributions. Self-interacting dark matter model. The halo mass $\log_{10} M_{200}$, the velocity-weighted cross section $\log_{10} \langle \sigma v \rangle / m$, and the dynamical mass $\log_{10} M_{\mathrm{dyn}}$ are derived parameters.}\label{fig:posteriors_SIDM}
\end{figure*}

\begin{figure*}
    \centering
    \includegraphics[scale=0.41]{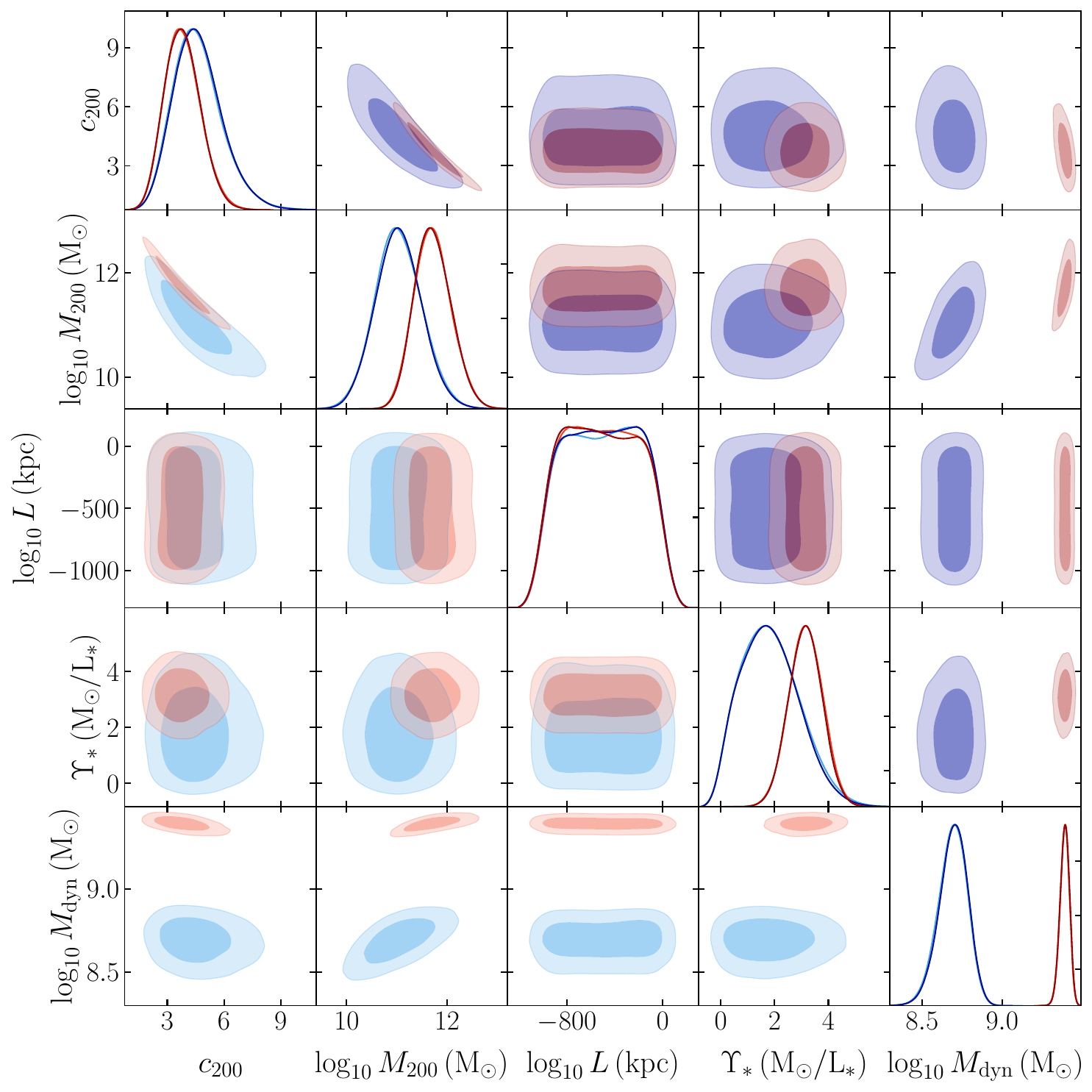} \\[0.2cm]
    ~~~~~~~~~~~\includegraphics[scale=0.58]{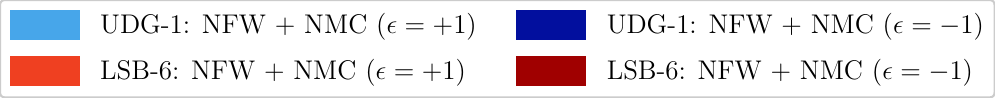} \\[0.6cm]
    \includegraphics[scale=0.41]{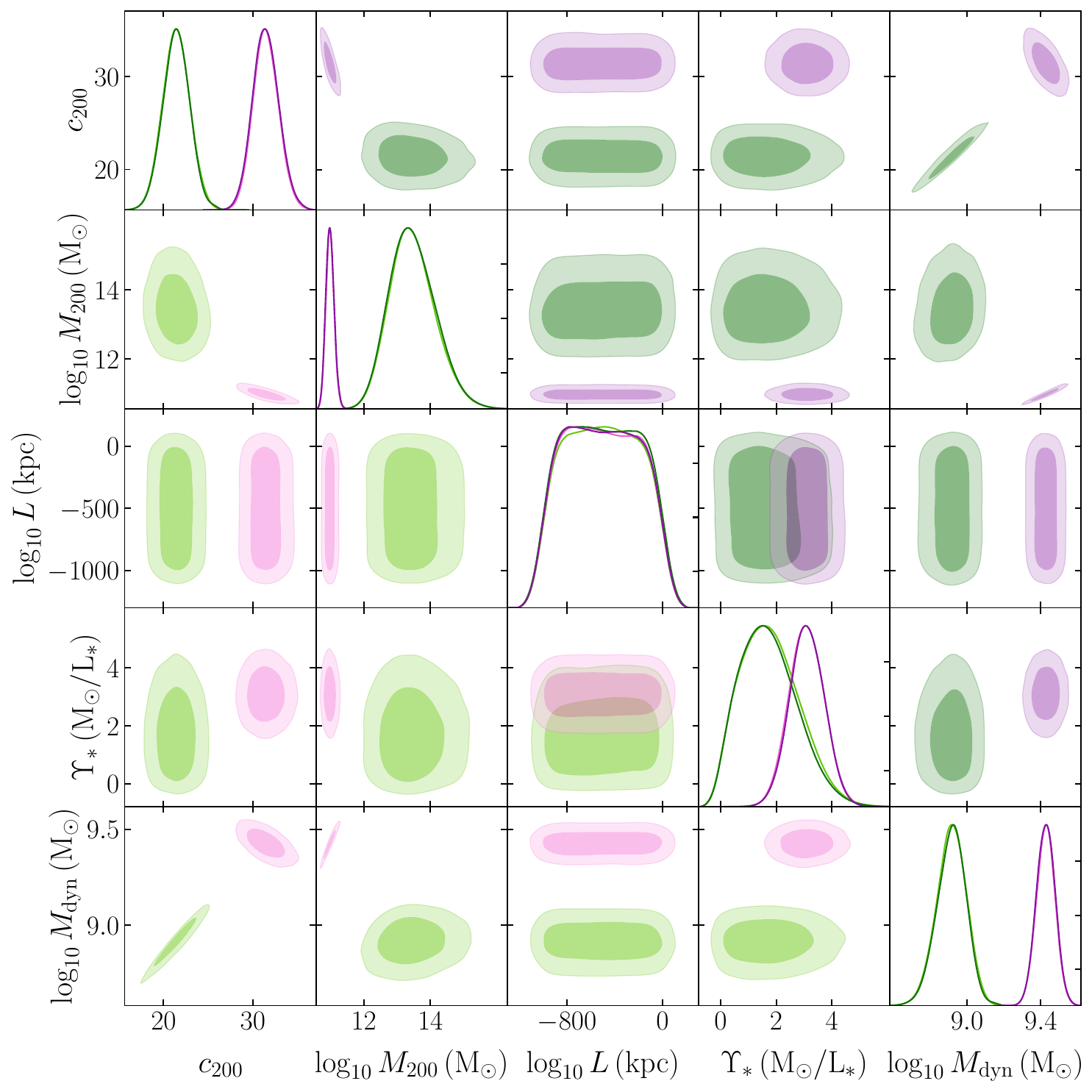} \\[0.2cm]
    ~~~~~~~~~~\includegraphics[scale=0.58]{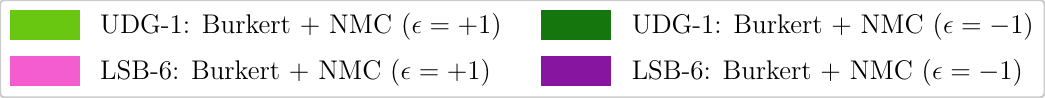}
    \caption{UDG-1 and LSB-6 posterior distributions. Non-minimally coupled dark matter models applied to the cuspy NFW (top) and the cored Burkert (bottom) profiles. The dynamical mass $\log_{10} M_{\mathrm{dyn}}$ is a derived parameter.}\label{fig:posteriors_NMC_NFW-Burkert}
\end{figure*}

\begin{figure*}
    \centering
    \includegraphics[scale=0.55]{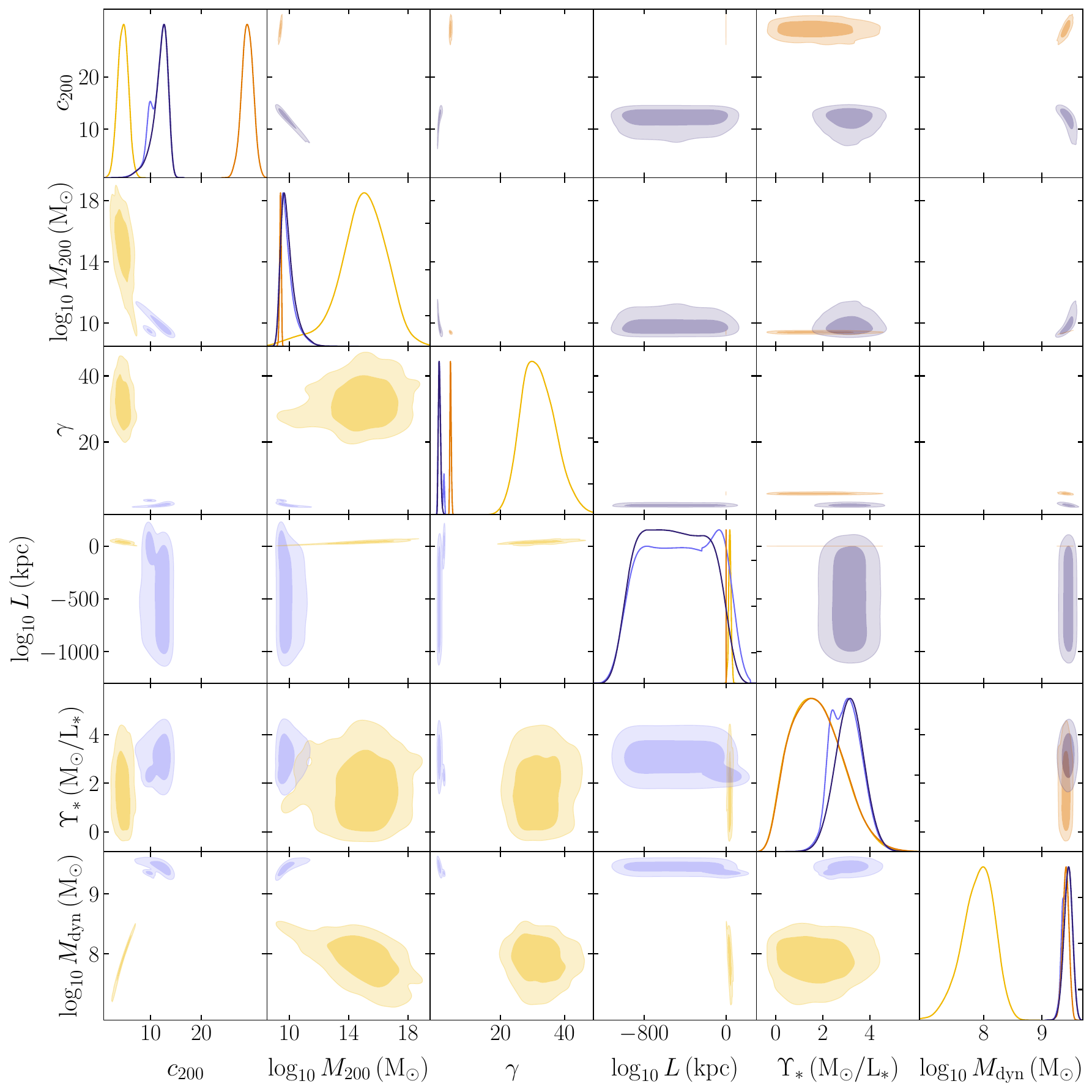} \\[0.4cm]
    ~~~~~~~~~~~~~\includegraphics[scale=0.65]{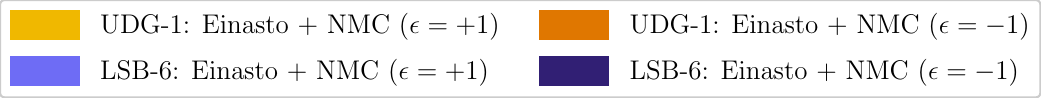}
    \caption{UDG-1 and LSB-6 posterior distributions. Non-minimally coupled dark matter model applied to the soft-cored Einasto profile. The dynamical mass $\log_{10} M_{\mathrm{dyn}}$ is a derived parameter.}\label{fig:posteriors_NMC_Einasto}
\end{figure*}

\clearpage
\stepcounter{section}   
\setcounter{figure}{0}  

\begin{figure*}[b]
    \centering
    \vspace{0.1cm}
    \includegraphics[scale=0.78]{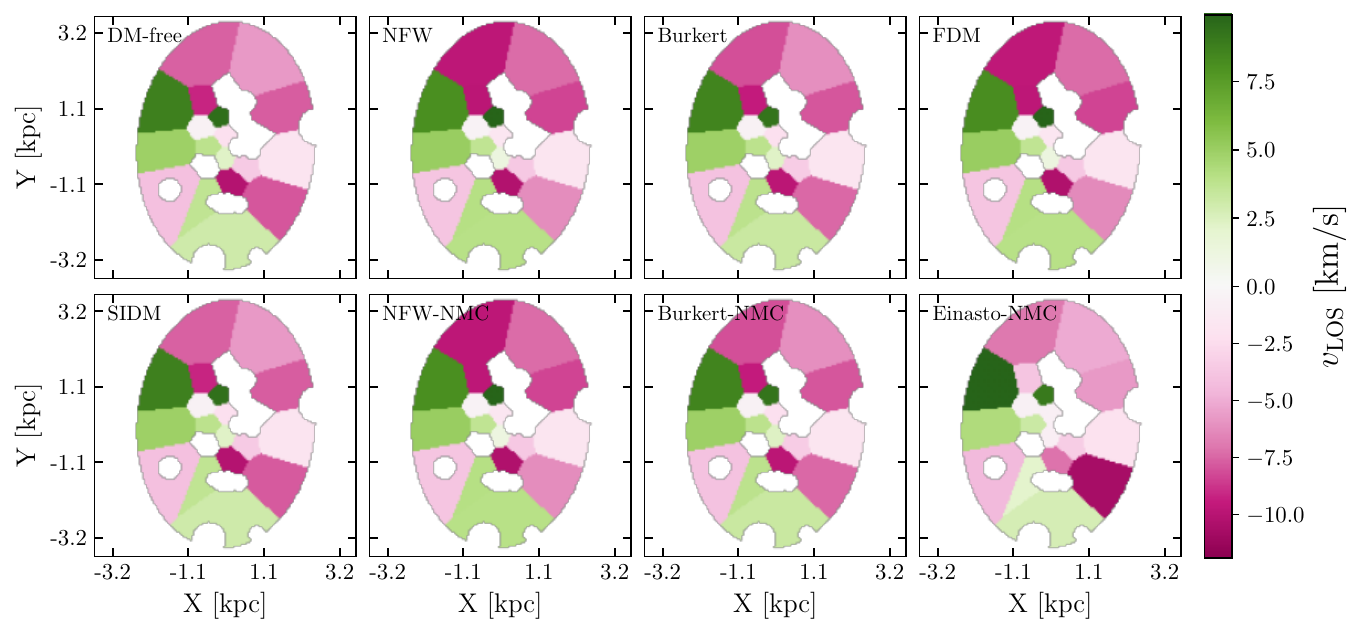}
    \caption{UDG-1 normalized residual maps. All quantities are computed using the median parameter values reported in Table~\ref{tab:UDG1}. The non-minimally coupled models are shown for the negative-coupling case ($\epsilon=-1$), while the fiducial Einasto model is presented in Sect.~\ref{sec:results}.}\label{fig:UDG1_residuals}
\end{figure*}

\begin{figure*}[b]
    \centering
    \vspace{0.5cm}
    \includegraphics[scale=0.78]{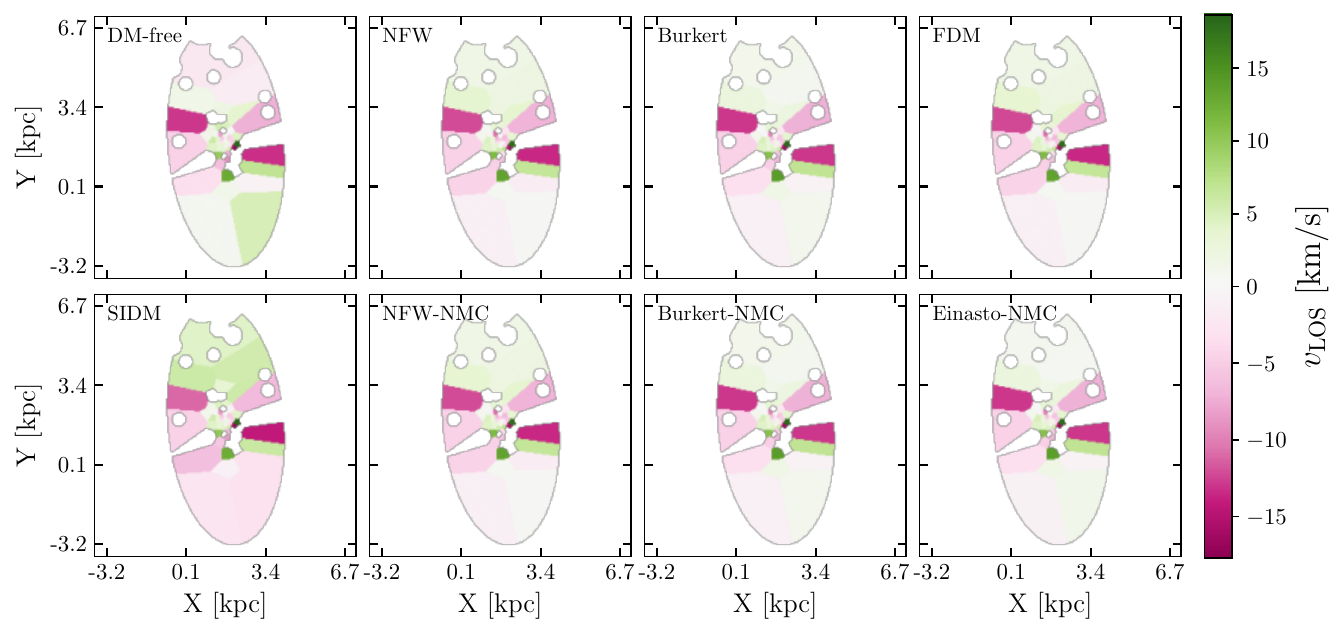}
    \caption{LSB-6 normalized residual maps. All quantities are computed using the median parameter values reported in Table~\ref{tab:LSB6}. The non-minimally coupled models are shown for the negative-coupling case ($\epsilon=-1$), while the fiducial Einasto model is presented in Sect.~\ref{sec:results}.}\label{fig:LSB6_residuals}
    \vspace{1cm}
\end{figure*}

\addtocounter{section}{-1}
\section{Residual maps}\label{AppendixC:residuals}

In this appendix we present the residual maps obtained from our Bayesian analysis of the kinematics of UDG-1 and LSB-6. The residual maps for the fiducial Einasto model are not shown, as they are already reported in Sec.~\ref{sec:results}.

\end{appendix}

\end{document}